\documentclass[fina2,3l,leqno]{siamltex}
\usepackage{amsmath}
\usepackage{amsfonts}
\usepackage{amssymb}
\usepackage{enumerate}
\usepackage{subfigure}
\usepackage{color}
\usepackage{graphicx}
\usepackage{verbatim}
\usepackage{url}
\usepackage{alltt}   
\usepackage{algorithm2e}
\usepackage{qtree}
\usepackage{mathrsfs}
\usepackage{eufrak}
\usepackage{yfonts}

\usepackage{simplemargins}
\setleftmargin{1in}
\setrightmargin{1in}
\settopmargin{1in}
\setbottommargin{1in}
    
\newcounter{saveeqn}%

\DeclareMathAlphabet{\mathpzc}{OT1}{pzc}{m}{it}

\newtheorem{theo}{Theorem}[section]
\newtheorem{lemm}{Lemma}[section]

\newtheorem{Defn}{Definition}[section]

\author{E.R. Swart\footnotemark[1], S.J. Gismondi\footnotemark[2],  N.R. Swart\footnotemark[3], C.E. Bell\footnotemark[4], A. Lee\footnotemark[8]}
\title{Deciding Graph non-Hamiltonicity via a Closure Algorithm} 
\begin{document}
\maketitle
\renewcommand{\thefootnote}{\fnsymbol{footnote}}

\footnotetext[5]{\textit{This work is dedicated to the late Linda Allen, friend, University of Guelph colleague and mentor.}}
\footnotetext[1]{\textit{Kelowna, British Columbia, Canada, Email: \textup{\nocorr \texttt{ted.swart@shaw.ca}}}}
\footnotetext[2]{\textit{University of Guelph, Canada, Email: \textup{\nocorr \texttt{gismondi@uoguelph.ca}}} (Corresponding Author)}
\footnotetext[3]{\textit{University of British Columbia Okanagan, Canada, Email: \textup{\nocorr \texttt{nicholas.swart@shaw.ca}}}}
\footnotetext[4]{\textit{Guelph, Ontario, Canada, Email: \textup{\nocorr \texttt{ca\_bell@rogers.com}}}}
\footnotetext[8]{\textit{University of Guelph, Canada, Email: \textup{\nocorr \texttt{alee15@mail.uoguelph.ca}}}}

\begin{abstract}We present a matching and LP based heuristic algorithm that decides graph non-Hamiltonicity. Each of the $n!$ Hamilton cycles in a complete directed graph on $n+1$ vertices corresponds with each of the $n!$ $n$-permutation matrices $P$, such that $p_{u,i}=1$ if and only if the $i^{th}$ arc in a cycle enters vertex $u$, starting and ending at vertex $n+1$. A graph instance ($G$) is initially coded as exclusion set $E$, whose members are pairs of components of $P$, $\{p_{u,i} ,p_{v,i+1}\}, i=1,n-1$, for each arc $(u,v)$ not in $G$. For each $\{p_{u,i} ,p_{v,i+1}\}\in E$, the set of $P$ satisfying $p_{u,i}=p_{v,i+1}=1$ correspond with a set of cycles not in $G$. Accounting for all arcs not in $G$, $E$ codes precisely the set of cycles not in $G$. A doubly stochastic-like $\mathcal{O}$($n^4$) formulation of the Hamilton cycle decision problem is then constructed. Each $\{p_{u,i} ,p_{v,j}\}$ is coded as variable $q_{u,i,v,j}$ such that the set of integer extrema is the set of all permutations. We model $G$ by setting each $q_{u,i,v,j}=0$ in correspondence with each $\{p_{u,i} ,p_{v,j}\}\in E$ such that for non-Hamiltonian $G$, \textit{integer solutions cannot exist}. We then recognize non-Hamiltonicity by iteratively deducing additional $q_{u,i,v,j}$ that can be set zero and expanding $E$ until the formulation becomes infeasible, in which case we recognize that no integer solutions exists i.e. $G$ is decided non-Hamiltonian. The algorithm first chooses any $\{p_{u,i} ,p_{v,j}\} \not \in E$ and sets $q_{u,i,v,j}=1$. \textit{As a relaxed LP}, if the formulation is infeasible, we deduce $q_{u,i,v,j}=0$ and $\{p_{u,i} ,p_{v,j}\}$ can be added to $E$. Then we choose another $\{p_{u,i} ,p_{v,j}\} \not \in E$ and start over. Otherwise, \textit{as a subset of matching problems} together with a subset of necessary conditions, if $q_{u,i,v,j}$ cannot participate in a match, we deduce $q_{u,i,v,j}=0$ and $\{p_{u,i} ,p_{v,j}\}$ can be added $E$. We again choose another $\{p_{u,i} ,p_{v,j}\} \not \in E$ and start over. Otherwise $q_{u,i,v,j}$ is undecided, and we exhaustively test all $\{p_{u,i} ,p_{v,j}\} \not \in E$. If $E$ becomes the set of all $\{p_{u,i} ,p_{v,j}\}$, $G$ is decided non-Hamiltonian.  Otherwise $G$ is undecided. We call this the Weak Closure Algorithm. Only non-Hamiltonian $G$ share this maximal property. Over 100 non-Hamiltonian graphs (10 through 104 vertices) and 2000 randomized 31 vertex non-Hamiltonian graphs are tested and correctly decided non-Hamiltonian. For Hamiltonian $G$, the complement of $E$ provides information about covers of matchings, perhaps useful in searching for cycles. We also present an example where the WCA fails to deduce any integral value for any $q_{u,i,v,j}$ i.e. $G$ is undecided.\end{abstract}


\thispagestyle{empty}


\markboth{Swart \& Gismondi \& Bell \& Swart}{Graph Non-Hamiltonicity via Closure} 

\section{Introduction}
\label{sec:SummaryIntro}\setcounter{equation}{-1}\setcounter{equation}{0}We present a theory, model, and an heuristic algorithm that shows how to decide problems in \textbf{NP} via models of their \textbf{coNP} counter parts, based upon the Birkhoff polytope. We first model the non-Hamilton cycle decision problem to create a relaxed linear programming formulation of the Hamilton cycle decision problem. We then present the $\mathcal{O}(n^8)$ \textit{Weak Closure Algorithm} (WCA) that deduces values of 0/1 variables via Boolean closure (in place of LP) and non-matching. Over 100 non-Hamiltonian graphs (10 through 104 vertices) (SJG, NRS \& AL),  and 2000 randomized 31 vertex non-Hamiltonian graphs (NRS) are tested and correctly decided non-Hamiltonian\footnotemark[7]. Graphs require no special treatment. No tested graphs failed that were not reported. We believe that the relaxed linear programming formulation models (the $Q$ matrix formulation) useful and possibly new kinds of insights/relationship between permutations, exploited by the WCA. We invite researchers to investigate and develop these ideas with us. Please contact the corresponding author for FORTRAN code and details about test graphs etc.\footnotetext[7]{\textit{NRS used Matlab to generate these 2000 graphs.}}\

The WCA can also be used to verify non-Hamiltonicity, given a correctly guessed set of variables as input to the WCA. See section \ref{sec:TermFeasWCA}. The WCA is easy to apply to 1) a model of the graph isomorphism decision problem (section \ref{sec:coNPmods}) partially answering a question in \cite{gismo13a} i.e. we show how to generate the input set sufficient to model its \textbf{coNP} counter model and 2) the subgraph isomorphism decision problem modelled in \cite{gismo08} i.e. we can create a model of the subgraph non-isomorphism decision problem, exclusion set $E$, as input to the WCA. This is expected, given that we model the \textbf{NP}-complete Hamilton Cycle Decision Problem. For these reasons we propose that problems in \textbf{coNP} be modelled and studied as compact formulations whose set of extrema are sets of permutations in correspondence with sets of non-solutions i.e. a unified approach based upon permutations. Information about what is not a solution might be used to create an input set that models \textbf{NP} counter problems (unrelated to the complexity of deciding an \textbf{NP} problem). It might also be convenient to modify the WCA to become more exhaustive/complex. So we comment that the WCA is generalizable/parallelized (as implemented by SJG). See section \ref{sec:GenWCA}.\

For special classes of graphs, we believe that the WCA can be developed as a technique that always decides graph non-Hamiltonicity e.g. snarks. We would first need to prove there exists sufficient polynomial time accessible information via $E$ (see section \ref{ssec:Justify1}) for \textit{all} instances of `YES' decisions to problems in \textbf{coNP}, and then prove that the WCA must always cause the corresponding relaxed formulation of the \textbf{NP} problem to become infeasible. If these classes of graphs exists, and these proofs were known, the of course feasible solutions imply the existence of an integer solution i.e. a `YES' decision for problems in \textbf{NP}. This is how we envision a highly practical use of the WCA.\

\subsection{Our Motivation to Study \textbf{coNP}-complete Decision Problems}
\label{sec:MotivationOur}\setcounter{equation}{-1}\setcounter{equation}{0}
During the 1990s, two of us (SJG and ERS) began modelling \textbf{NP} decision problems as compact relaxed linear programming formulations of integer programs such that integer extrema are permutations that correspond with solutions i.e. our models are based upon the Birkhoff polytope. A common idea at that time (among the \textbf{P}=\textbf{NP} proponents) was to search for a compact linear programming formulation of an \textbf{NP}-complete problem, infeasible if and only if there exists no integer solutions i.e. implying that \textbf{P}=\textbf{NP} \cite{cowo08,gismo03,j87,j92}. Unsuccessful (and maybe impossible \cite{fmptw12}), these ideas were applied to their \textbf{coNP} counter models, compact formulations of the union of sets of permutations using projection and lifting techniques, each permutation in correspondence with each non-solution. These ideas were formalized in a graduate student thesis, leading to a series of small results \cite{dg08,gismo03,gismo08,gismo13a,gismo01,gs04} occasionally presented at conferences\footnotemark[6]. Interestingly for some instances of non-Hamiltonian graphs, it's easy (polynomial time) to deduce that the set of all non-solutions is the set of all permutations i.e. easy to decide non-Hamiltonicity. These non-Hamiltonian graphs (and their models) share a property that we exploit via LP and it's not necessary to search an intractable space of solutions. This is common idea, analogous to how Michael Sipser describes an approach to `what is not prime' in \cite{sipvid}. \footnotetext[6]{\textit{Presentations at: the 22$^{nd}$ Southeastern International Conference on Combinatorics, Graph Theory, and Computing in Baton Rouge; University of Manitoba 1992; INFORMS 2009; 8FCC 2010; ICGT 2014.}}\

\subsection{General Motivation to Study \textbf{coNP}-complete Decision Problems}
\label{sec:Motivation}\setcounter{equation}{-1}\setcounter{equation}{0}
Research related to the infamous \textbf{NP}-hard Traveling Salesman Problem (TSP) spans many years and perhaps thousands of researchers \cite{aa03,aa09,abcc01,abcc03,abcc06,bs96,book92,j87,j92,llks85,sip92}, to mention just a few. TSP polytope facet finding studies date back to at least the 1950's with Heller, Kuhn, Norman and Robacker as cited in \cite{schrij03}. At that time, the concept of \textbf{NP}-completeness hadn't been formalized although the TSP problem was suspected to be very complicated, noted by Flood also cited in \cite{schrij03}.  Studies in complexity theory, and the field itself are well developed due in large part to the high profile of the \textbf{P} versus \textbf{NP} conundrum, one of six remaining Millennium problems \cite{coo06}, not to mention applications to communications, security - cryptography in particular \cite{diffie1976}.  Quantum computation techniques have generated further important contributions \cite{fmptw12,shor1994,shor1997}, preceded by the development of quantum complexity theory \cite{deutsch1985}, now a mainstream research area.\

From 1990 - 2015, researchers have regularly proposed proofs that resolve the \textbf{P} versus \textbf{NP} conundrum \cite{Fort2009,Fort2013,li12,li12a,wo12}. It seems that such frequent and prolonged interest should generate advances. But there are simply very few, perhaps none to be had, or maybe we lack insight. Regardless, much less effort appears to be spent on studying \textbf{NP} versus \textbf{coNP}, other than as a byproduct of \textbf{NP}-completeness. Some work on \textbf{coNP}-completeness, MNH \& hypohamiltonian graphs can be found in \cite{eh83,fjp12,g89,Jamrozik1982229, ro03,ro04,ls98,we12,jz77}. Noting that \textbf{NP} $\ne$ \textbf{coNP} $\Rightarrow$ \textbf{P} $\ne$ \textbf{NP}, we propose that researchers should study \textbf{NP} versus \textbf{coNP} via the creation and study of models and algorithms that \textit{decide} YES to \textbf{coNP}-complete problems. Is there a \textit{good} way to 1) access information from models of \textbf{coNP}-complete problems so that we can make \textit{good} use of this information to solve \textbf{NP}-complete problems? (\textbf{P} versus \textbf{NP}), and 2) make use of information we access from models of \textbf{coNP}-complete problems together with polynomial amounts of additional information to verify correctly guessed `YES' instances of \textbf{coNP}-complete problems? (\textbf{coNP} versus \textbf{NP}).

\subsection{Introduction to a Model of the Hamilton Cycle Decision Problem}
\label{sec:IntroModel}\setcounter{equation}{-1}\setcounter{equation}{0}
Let $G$ be a simple, strongly connected and directed $n+1$ vertex graph, neither empty nor complete where undirected edges are regarded as pairs of counter directed arcs. A Hamilton cycle in $G$ (cycle) is a directed circuit containing all vertices in $G$ and $n+1$ arcs in the arc set of $G$. $G$ is (non-)\textit{Hamiltonian} if and only if there exists (no) a cycle in $G$.  The problem of deciding $G$ (non-)Hamiltonian is called `The (non-)Hamilton cycle decision problem' and is (\textbf{coNP}) \textbf{NP}-complete. \cite{gjt76}. \

Cycles are permutations of vertex labels of $G$. One way (of many possible ways, see section \ref{sec:PracticalWCA3}) to model this idea is to assign each $n+1$ cycle to be in bijective correspondence with each $n$-permutation matrix $P$ such that $p_{u,i}$=1 if and only if the $i^{th}$ arc in a cycle enters vertex $u$, starting and ending at vertex $n+1$.\

Let each $P$ of the set of $n!$ $P$ matrices be coded within an $n^2$x$n^2$ matrix $Q$ such that the block structure of each $Q$ is the component structure of each $P$. See an example in Figure \ref{samplemat} ($n$=3). The general form of $Q$ is shown in Figure \ref{Qmat} ($n$=4) where components of $Q$ ($q$ variables) are written as $q_{u,i,v,j}$, referring to component $(v,j)$ in block $(u,i)$; and components of $P$ ($p$ variables) are written as $p_{u,i}$ referring to component $(u,i)$ in $P$. Note that when $j<i$, $q_{u,i,v,j}$ is written as $q_{v,j,u,i}$, a convenience based upon interpretation of these variables as applied to a model of the Hamilton cycle decision problem.\  
  
System 1 is a compact system of linear equations with fractional and integer extrema\footnotemark[8]. Each integer extreme is a permutation matrix $P$ that uniquely extends to 0/1 $q$ variables i.e. there exist precisely $n!$ unique 0/1 assignments of $q$ (or $p$) variables such that $\frac{n(n-1)}{2}$ (or $n$) $q$ (or $p$) variables set at unit level that cause by linear dependencies (System 1), an assignment of $n$ (or $\frac{n(n-1)}{2}$)  $p$ ( or $q$) variables at unit level \cite{gismo08}.\
\footnotetext[8]{\textit{See section \ref{ssec:Justify} re: likely non-existence of any compact formulation that models any \textbf{NP}-complete problem.}}
  
\begin{figure}[h!]
\begin{center}
\small
$\left(\begin{array}{ccc|ccc|ccc}
           0 & 0 & 0 & 0 & 0 & 0 & 0 & 0 & 1 \\
           0 & 0 & 0 & 0 & 0 & 0 & 1 & 0 & 0 \\
           0 & 0 & 0 & 0 & 0 & 0 & 0 & 1 & 0 \
\\\hline           
           0 & 0 & 1 & 0 & 0 & 0 & 0 & 0 & 0 \\
           1 & 0 & 0 & 0 & 0 & 0 & 0 & 0 & 0 \\
           0 & 1 & 0 & 0 & 0 & 0 & 0 & 0 & 0 \
\\\hline
           0 & 0 & 0 & 0 & 0 & 1 & 0 & 0 & 0 \\
           0 & 0 & 0 & 1 & 0 & 0 & 0 & 0 & 0 \\
           0 & 0 & 0 & 0 & 1 & 0 & 0 & 0 & 0 \\
\end{array}\right)$
\normalsize
\caption{An example of the block structure of $Q$ defined by the component structure of $P$, $n=3$.}
\label{samplemat}
\end{center}
\end{figure}

\begin{figure}[h!]
\begin{center}
\tiny
$\left(\begin{array}{cccc|cccc|cccc|cccc} 
p_{11} & 0 & 0 & 0 & 0 & p_{12} & 0 & 0 & 0 & 0 & p_{13} & 0 & 0 & 0 & 0 & p_{14} \\
0 & q_{1122} & q_{1123} & q_{1124} & q_{2112}& 0 & q_{1223} & q_{1224} & q_{2113} & q_{2213} & 0 & q_{1324} & q_{2114} & q_{2214} & q_{2314} & 0 \\ 
0 & q_{1132} & q_{1133} & q_{1134} & q_{3112}& 0 & q_{1233} & q_{1234} & q_{3113} & q_{3213} & 0 & q_{1334} & q_{3114} & q_{3214} & q_{3314} & 0 \\ 
0 & q_{1142} & q_{1143} & q_{1144} & q_{4112}& 0 & q_{1243} & q_{1244} & q_{4113} & q_{4213} & 0 & q_{1344} & q_{4114} & q_{4214} & q_{4314} & 0 \\ 
\\\hline
0 & q_{2112} & q_{2113} & q_{2114} & q_{1122} & 0 & q_{2213} & q_{2214} & q_{1123} & q_{1223} & 0 & q_{2314} & q_{1124} & q_{1224} & q_{1324} & 0 \\
p_{21} & 0 & 0 & 0 & 0 & p_{22} & 0 & 0 & 0 & 0 & p_{23} & 0 & 0 & 0 & 0 & p_{24} \\
0 & q_{2132} & q_{2133} & q_{2134} & q_{3122}& 0 & q_{2233} & q_{2234} & q_{3123} & q_{3223} & 0 & q_{2334} & q_{3124} & q_{3224} & q_{3324} & 0 \\ 
0 & q_{2142} & q_{2143} & q_{2144} & q_{4122}& 0 & q_{2243} & q_{2244} & q_{4123} & q_{4223} & 0 & q_{2344} & q_{4124} & q_{4224} & q_{4324} & 0 \\ 

\\\hline
0 & q_{3112} & q_{3113} & q_{3114} & q_{1132} & 0 & q_{3213} & q_{3214} & q_{1133} & q_{1233} & 0 & q_{3314} & q_{1134} & q_{1234} & q_{1334} & 0 \\
0 & q_{3122} & q_{3123} & q_{3124} & q_{2132} & 0 & q_{3223} & q_{3224} & q_{2133} & q_{2233} & 0 & q_{3324} & q_{2134} & q_{2234} & q_{2334} & 0 \\
p_{31} & 0 & 0 & 0 & 0 & p_{32} & 0 & 0 & 0 & 0 & p_{33} & 0 & 0 & 0 & 0 & p_{34} \\
0 & q_{3142} & q_{3143} & q_{3144} & q_{4132}& 0 & q_{3243} & q_{3244} & q_{4133} & q_{4233} & 0 & q_{3344} & q_{4134} & q_{4234} & q_{4334} & 0 \\
\\\hline
0 & q_{4112} & q_{4113} & q_{4114} & q_{1142} & 0 & q_{4213} & q_{4214} & q_{1143} & q_{1243} & 0 & q_{4314} & q_{1144} & q_{1244} & q_{1344} & 0 \\
0 & q_{4122} & q_{4123} & q_{4124} & q_{2142} & 0 & q_{4223} & q_{4224} & q_{2143} & q_{2243} & 0 & q_{4324} & q_{2144} & q_{2244} & q_{2344} & 0 \\
0 & q_{4132} & q_{4133} & q_{4134} & q_{3142} & 0 & q_{4233} & q_{4234} & q_{3143} & q_{3243} & 0 & q_{4334} & q_{3144} & q_{3244} & q_{3344} & 0 \\
p_{41} & 0 & 0 & 0 & 0 & p_{42} & 0 & 0 & 0 & 0 & p_{43} & 0 & 0 & 0 & 0 & p_{44} \\
\end{array}\right)$
\normalsize
\caption{General Form of $Q$ Matrix, $n=4$.}
\label{Qmat}
\end{center}
\end{figure}

\textbf{System 1:}\
     $\,\,\sum_{i}p_{u,i}=1, u=1,2,...,n$\

     $\,\,\,\,\,\,\,\,\,\,\,\,\,\,\,\,\,\,\,\,\,\,\,\,\,\,\,\,\,\,\,\,\,\sum_{u}p_{u,i}=1, i=1,2,...,n$\
      
     $\,\,\,\,\,\,\,\,\,\,\,\,\,\,\,\,\,\,\,\,\,\,\,\,\,\,\,\,\,\,\,\,\,For\,\,all\,\,u,i=1,2,...,n$
      
     $\,\,\,\,\,\,\,\,\,\,\,\,\,\,\,\,\,\,\,\,\,\,\,\,\,\,\,\,\,\,\,\,\,\,\,\,\,\,\,\,\sum_{j\neq i}{^1}q_{u,i,v,j} = p_{u,i}, v=1,2,...,n$, $v\neq u$.\

     $\,\,\,\,\,\,\,\,\,\,\,\,\,\,\,\,\,\,\,\,\,\,\,\,\,\,\,\,\,\,\,\,\,\,\,\,\,\,\,\,\sum_{v\neq u}{^1}q_{u,i,v,j} = p_{u,i}, j=1,2,...,n$, $j\neq i$.\       
\\
 
      $\,\,\,\,\,\,\,\,\,\,\,\,\,\,\,\,\,\,\,\,\,\,\,\,\,\,\,\,\,\,\,\,\,{^1}$\textbf{Note:} For $i>j$, write $q_{v,j,u,i}$ in place of $q_{u,i,v,j}$.\\

For $q_{u,i,v,j}$=0, precisely those $P$ for which $p_{u,i}=p_{v,j}=1$ are infeasible wrt System 1. Consider System 2 below, where a set of $q$ variables have been set zero (via set $E$) corresponding with elements of $E$, pairs of $p$ variables that cannot both be at unit level in any 0/1 solution. We code for sets of cycles excluded from $G$ as follows. For directed graph $G$, we examine all arcs $(u,v)$ not in $G$, accounting for every sequence position of every arc in every cycle not in $G$. $E$ is sufficient since $P$ corresponds with a cycle not in $G$ if and only if the cycle makes use of at least one arc not in $G$ i.e. requiring that $p_{u,i}=p_{v,i+1}=1$. We can exclude these cycles by placing 0's in positions $(u,i,v,i+1)$ of the $Q$ matrix in Figure \ref{Qmat}. System 2 is a relaxed model of the Hamilton cycle decision problem. Its set of integer extrema correspond with cycles in $G$ if and only if for each extreme $P$, the set of $\{p_{u,i},p_{v,j}\}$  $\not\in$ $E$ satisfying $p_{u,i}=p_{v,j}=1$ define $P$. $G$ is non-Hamiltonian if and only if System 2 is either infeasible or has only fractional extrema.\footnotemark[9]
\footnotetext[9]{\textit{This is how we use information from a model of the non-Hamilton cycle decision problem to create a relaxed linear programming formulation of the Hamilton cycle decision problem.}}\

It's also possible to make deductions from $E$ in ways that the WCA cannot, in advance of the WCA for application in the WCA (implemented in our code). Using Dijkstras algorithm, the shortest path between vertices $u$ and $v$ is of length $k$, and thus paths of length $k-1,k-2,....1$ do not exist i.e. $\{p_{u,i},p_{v,i+k-1}\}, \{p_{u,i},p_{v,i+k-2}\},...,\{p_{u,i},p_{v,i+1}\}$ can be added to $E$ at the outset. We gain this new information (the shortest path matrix) in polynomial time, therefore called polynomial time accessible information.\
\\

\textbf{System 2:}\
     $\,\,\sum_{i}p_{u,i}=1, u=1,2,...,n$\

     $\,\,\,\,\,\,\,\,\,\,\,\,\,\,\,\,\,\,\,\,\,\,\,\,\,\,\,\,\,\,\,\,\,\sum_{u}p_{u,i}=1, i=1,2,...,n$\
      
     $\,\,\,\,\,\,\,\,\,\,\,\,\,\,\,\,\,\,\,\,\,\,\,\,\,\,\,\,\,\,\,\,\,For\,\,all\,\,u,i=1,2,...,n$
      
     $\,\,\,\,\,\,\,\,\,\,\,\,\,\,\,\,\,\,\,\,\,\,\,\,\,\,\,\,\,\,\,\,\,\,\,\,\,\,\,\,\sum_{j\neq i}{^1}q_{u,i,v,j} = p_{u,i}, v=1,2,...,n$, $v\neq u$.\

     $\,\,\,\,\,\,\,\,\,\,\,\,\,\,\,\,\,\,\,\,\,\,\,\,\,\,\,\,\,\,\,\,\,\,\,\,\,\,\,\,\sum_{v\neq u}{^1}q_{u,i,v,j} = p_{u,i}, j=1,2,...,n$, $j\neq i$.\       

     $\,\,\,\,\,\,\,\,\,\,\,\,\,\,\,\,\,\,\,\,\,\,\,\,\,\,\,\,\,\,\,\,\,For\,\,each\,\, \{p_{u,i},p_{v,j}\}\in E\,\,assign\,\,q_{u,i,v,j}=0$.\
           
     $\,\,\,\,\,\,\,\,\,\,\,\,\,\,\,\,\,\,\,\,\,\,\,\,\,\,\,\,\,\,\,\,\,p_{u,i},q_{u,i,v,j}\geq 0$.
\\  
  
      $\,\,\,\,\,\,\,\,\,\,\,\,\,\,\,\,\,\,\,\,\,\,\,\,\,\,\,\,\,\,\,\,\,{^1}$\textbf{Note:} For $i>j$, write $q_{v,j,u,i}$ in place of $q_{u,i,v,j}$.\

\subsection{Introduction to the WCA}
\label{sec:IntroAlgo}\setcounter{equation}{-1}\setcounter{equation}{0}
We assume that System 2 has an integer solution $P$ and we seek to deduce infeasibility in the case of non-Hamiltonian $G$.  We systematically choose and then overlay pairs of blocks in $Q$ consistent with a subset of the component structure of an assumed $P$ and then model $P$'s common non-zero $p$ variables as arcs of an undirected bipartite graph (row and column vertices). This is equivalent to assuming a necessary condition for both blocks to participate in a match i.e. that a $q$ variable might be allowed to attain unit level in a solution to System 2. So we effectively set a variable at unit level, and if the LP is infeasible (although we use Boolean closure for implementation) or if the LP is feasible but there is no match, we deduce that the variable can be set zero. In this way, the WCA iteratively deduces additional $q$ variables to be set zero and corresponding pairs of $p$ variables to be added to $E$. Input for the WCA is therefore $E$. Output from the WCA is $\tilde{E}$, called a \textit{weak closure set} of $E$, defined temporarily (formalized in section \ref{ssec:PR}) below. Formal presentation of the WCA follows in section \ref{sec:MaxCoverAlgo}.\ 

\begin{Defn}Define $\tilde{E}$ = $E$ $\cup$ $\{$all pairs of $p$ variables whose corresponding $q$ variables are deduced zero via the WCA$\}$.\end{Defn}\

We recognize non-Hamiltonicity if $\tilde{E}$ is the set of all pairs of $p$ variables. Pre-emptively we might also test and confirm infeasibility of System 2 (but replace $E$ with $\tilde{E}$) and we say that the WCA \textit{decides non-Hamiltonicity} (temporarily ignoring a feasible System 2 for which there exists no integer solutions).\

Our overall approach therefore is to model a \textbf{coNP}-complete decision problem via $E$. We then create its corresponding \textbf{NP}-complete LP formulation and implement the WCA to 1) verify non-Hamiltonicity for correctly guessed non-Hamiltonian $G$ given correctly guessed and verified $E$, and 2) decide non-Hamiltonicity for $G$ using a polynomial time deterministically computed $E$. In the former case, this amounts to the study of \textbf{NP} versus \textbf{coNP}, and in the latter case, study of  \textbf{P} versus \textbf{NP}. Regarding (1), verifying non-Hamiltonicity is about verifying that System 2 admits no integer solution. $E$ is given and or/guessed in a `known way' sufficient for the WCA to verify non-Hamiltonicity i.e. that $G$ is non-Hamiltonian, requires knowledge that each element of $E$ is `a priori' known to be a pair of $p$ variables whose corresponding $q$ variable can be set zero. Regarding (2), deciding non-Hamiltonicity is about deciding that System 2 admits no integer solution. We compute $E$ and we implement the WCA as a tool to help close $E$ so that we can recognize non-Hamiltonicity. \textit{But in both cases, when the WCA fails to close $E$, we cannot decide if either $G$ is Hamiltonian i.e. recognizing that $\tilde{E}$ cannot  be the set of all pairs of $p$ variables, or that $G$ is non-Hamiltonian and that $E$ is not sufficient and/or the WCA is unable to close $E$ to be the set of all pairs of $p$ variables.} This comment relates to properties of $E$/$G$, i.e. the existence of classes of graphs with the property that the WCA \textit{always} closes $E$.\

\section{Towards Development of a Theory}
\label{sec:TowardTheory}\setcounter{equation}{-1}\setcounter{equation}{0}
As an overall approach in the context of what's known and what not to do, we first show how the ideas proposed in this paper at first glance, appear to avoid some of the known `no-go' avenues and perhaps have some merit toward making headway in the study of these hard problems. Definitions and a proposed closure theory follows in sections \ref{ssec:TE} and \ref{sec:EbarTheory}.\

\subsection{About What's Known and What Not To Do}
\label{ssec:Justify}\setcounter{equation}{-1}\setcounter{equation}{0}\
Recall that the Hamilton cycle decision problem is the TSP with equally weighted arcs. Rather than view the TSP polytope as the convex hull of cycles, it's more insightful to appreciate its complicated and beautiful structure regarding how it models `YES' decisions.  A set of graph instances that contain the same set of cycles corresponds with the same $d$-supporting hyperplane of the TSP polytope, defined by the affine hull of these cycles. There exist $\mathcal{O}(n!)$ isomorphic $d$-supporting hyperplanes (up to permutation of vertex labels) for each set of these instances, and there exist $\mathcal{O}(2^{n^2})$ graph instances whose affine hull of cycles is dimension $d$, $d=0,1,2,...,D=dim(TSP\,\,polytope)-1$. The complete set of `YES' decisions are encoded by the TSP polytope as the set of supporting hyperplanes constructed by permutation of every combination of cycles allowed in every Hamiltonian graph instance. This \textit{is} \textbf{NP}-completeness, that `the convex hull of cycles' yields intricate and complicated consequences i.e. that the TSP polytope encodes intractable `amounts' of information (every single asymmetric special instance, through to every instance of every isomorph of every graph). The set of $D$-facets (facets) of the TSP polytope are unknown, their number is unknown, nor can an arbitrary inequality be verified in polynomial time as inducing a facet of the TSP polytope.\

In May of 2012, it was proven that no compact formulation of the TSP polytope exists \cite{fmptw12} (For more background on the TSP, see \cite{abcc01,llks85}.).  From \cite{fmptw12}, \textit{`We solve a 20-year old problem posed by Yannakakis and prove that there exists no compact LP whose associated polytope projects to the traveling salesman polytope, even if the LP is not required to be symmetric. Moreover, we prove that this holds also for the cut polytope and the stable set polytope. These results were discovered through a new connection that we make between one-way quantum communication protocols and semidefinite programming reformulations of LPs.'}. This result is significant having resolved a 20-year old problem, but not surprising, discussed above.\

Proponents of \textbf{P=NP} are likely indifferent to this result since they don't necessarily invoke (extended) formulations of the TSP polytope in attempted proofs e.g. Swart \cite{wo12}. Lipton also notes this in his blog, in the spirit of not yet dismissing the existence of compact formulations of \textbf{NP}-complete problems. He also comments \textit{`Then we need to see how far the logic of this paper \cite{fmptw12} (and a FOCS 2012 followup noted on the blog of my Tech colleague Pokutta) extends to alternative formulations.'}. What exactly \textit{is} an alternative formulation? The usual thought is to imagine a compact LP that models an \textbf{NP}-complete problem that is feasible if and only if the decision for a problem instance is `YES' (no need to illustrate a solution). These ideas are well known, at least since LP was shown to be in \textbf{P} \cite{j87,j92}. But LP is also \textbf{P}-complete.  Thus \textbf{P} $=$ \textbf{NP} if and only if there exists such a formulation (no need to illustrate a formulation). This is common knowledge, indirectly noted by another blogger \cite{li12} \textit{`... as linear programming is \textbf{P}-complete, proving in full generality that such a problem cannot be solved by linear programming is equivalent to proving \textbf{P} $\ne$ \textbf{NP}.'}.\

In conclusion, together with the comment that researchers generally agree that new fundamental insights are needed to study these problems \cite{li12a}, it seems reasonable to study the non-Hamilton cycle decision problem free of any polytope that can be suggested as isomorphic to the TSP polytope. Maybe we could model `tractable pieces of TSP polytope isomorphs' and then the algorithm becomes the focus of the study. Regardless, the entire approach must be reconciled and related to the existence or non-existence of a compact LP since LP is \textbf{P}-complete. What we propose amounts to investigating \textbf{coNP}-completeness using a model and algorithm intent on `staving off' intractabilities while shrinking the feasible set of solutions.\ 

\subsection{Force Infeasibility and Start-Stay-Stop Tractable}
\label{ssec:Justify1}\setcounter{equation}{-1}\setcounter{equation}{0}\
We comment that the success of the WCA depends upon deciding infeasibility of System 2 i.e. System 1 together with a subset of $q$ variables set at zero level. Either all pairs of $p$ variables are added to $\tilde{E}$ via the WCA, in which case System 2 is infeasible, or we explicitly monitor the feasibility of System 2 during implementation of the WCA. We wonder about the complexity of computing or verifying sufficient information about $E$ so as to efficiently `collapse' the feasible region of System 2 via the WCA. Hence we define and discuss the feasible region of Systems 1 \& 2.\

\begin{Defn}Define polytope $\mathbb{Q}$ as the solution set of System 1.\end{Defn}\

Noting `what not to do' from above, we now show how we develop the model and algorithm to respect some of these  boundaries. We first comment that cycles are permutations of vertex labels and that combinatorial decision problems are all about understanding the set of permutations. So it makes sense to map cycles to permutations and work entirely with permutations. Given the Birkhoff-von Neumann theorem and matching algorithm, we combine these ideas attempting to force / decide infeasibility of System 2.\

We first describe in more detail, how Birkhoff-like polytope $\mathbb{Q}$ is unlike the TSP polytope in that it's tractable yet perhaps still useful. Polytope $\mathbb{Q}$ was originally conceived by Ted Swart in 1990. Its set of 0/1 extrema are in bijective correspondence with extrema of the Birkhoff polytope a.k.a. cycles of a complete directed graph. They span the same affine subspace as does $aff$($\mathbb{Q}$). Thus $\mathbb{Q}$'s set of fractional extrema are affine combinations of its 0/1 extrema, and they project interior to the Birkhoff polytope i.e. onto images of convex combinations of integer extrema. The image of $\mathbb{Q}$ \textit{is} the Birkhoff polytope. As an aside, we also note that $\mathbb{Q}$ is the second member in a sequence of relaxed formulations of polyhedra whose limiting polytope is $\mathcal{Q}_{0/1}$ in \cite{dg08}, the convex hull of integer extrema of $\mathbb{Q}$. See \cite{gismo08,gismo13a} for related work. We also note that by restricting the feasible region to facets of $\mathbb{Q}$, we do not introduce additional extrema, unlike cutting plane methods etc.  We confirmed that $q_{u,i,v,j}$=0 are facets of $\mathbb{Q}$ (for $n$=4, 5) \cite{dg08} and this led to the creation of $E$.\

Elements of $E$ seed the `sequence of steps' referenced above, and how we go about `shrinking the feasible region' is described generally below. These initial set of constraints, together with the WCA dynamically generate more constraints i.e. they help to identify $q$ variables that can be set zero and cause more $q$ variables to be set zero. The algorithm we present implements LP techniques, testing individual $q$ variables to be set zero (below). Our FORTRAN code implements Boolean closure (speedier).\

We also bolster $E$, to create an $\mathcal{O}(n^4)$ exclusion set that might help speed infeasibility.  Construction of $E$ is formally presented in section \ref{sec:ExclusionSet} via Algorithm \ref{alg:ExcSet}. Given $G$, Dijkstra's algorithm is invoked to compute minimal paths of length $m$ between pairs of vertices implying that no paths of length less than $m$ exist. These paths are subpaths of Hamilton cycles not in $G$, coded as disallowed subsequences of vertex permutations (cycles). Corresponding variables are set zero, interpreted as: Hamilton cycles in $G$ cannot use subpaths corresponding to these variables. We note how this `should a.k.a. face value'  help cause exclusion of sufficient numbers of permutations so as to decide infeasibility. This is an example of adding polynomial time accessible information to $E$  i.e. recall this term from section \ref{sec:SummaryIntro}.\

The idea of `staving off' intractabilities from above is the design intent of the WCA. For non-Hamiltonian $G$, we try to draw conclusions by assuming the existence of an integer solution for which a $q$ variable attains unit level and we hope to recognize a contradiction, maintaining `polynomiality'. Weak closure is a term we use to describe a process that assumes the existence of a solution to a mixed 0/1 LP (System 2) with a particular variable set at zero or one, and 1) after application of any/all techniques that can be thought of e.g. matching, Boolean closure or LP, deduces a contradiction in order to conclude that the opposite of the assumption is true, and 2) assigns the variable accordingly which may or may not imply integral assignments to even more variables.  For example, suppose a particular $p$ or $q$ variable is deduced to be set one. Then the complementary row and column variables are deduced (and can be set) to be zero. For $q$ variables, this has implication in two blocks. In the context of polytope $\mathbb{Q}$, by setting sufficient numbers of variables in System 2 to zero, it's also possible to deduce that some $p$ variables must also be zero, and this can result in assigning a whole block of $q$ variables to be zero. Most generally, $p$ and $q$ variables can be deduced to be zero as follows. We first assume that an integer solution exists, and we choose a variable to maximize to unit level. If the variable cannot attain unit level, say it's maximum is 0.9 (LP), then it's set to zero level, perhaps excluding many integer and fractional solutions. Likewise if a variable cannot attain zero level, it's set to unit level, and by double stochastity this implies that many more variables are again set to zero level.\

In summary then, deducing those variables to be set zero begins to characterize every possible 0/1 solution, all of which lie in intersecting sets of facets of $\mathbb{Q}$. Applied to all $\mathcal{O}(n^4)$ variables as iterations progress, and if it's possible to set sufficient numbers of $q$ variables to zero, we hope to decide either infeasibility or the existence of an integer extreme. For non-Hamiltonian $G$, there are no integer extrema and so we hope for a quick collapse of $\mathbb{Q}$.\
    
\subsection{Formal Terminology and Exclusion Set $E$}\label{ssec:TE}Recall that all possible $n!$ cycles in $G$ start and end at vertex $n+1$.\

\textit{About arcs, paths and cycles not in $G$:} The term `arc (not) in $G$' means an arc (not) in the arc set of $G$, and this suggests other terms like `path(s) (not) in $G$' and `cycle(s) (not) in $G$. That is, a path or cycle not in $G$ requires that \textit{at least one arc} in the path or cycle not be in $G$, i.e. paths and cycles in $G$ require that \textit{all arcs} in these paths and cycles be in $G$. Consider now the set of all cycles in $G$. The complement of this set is the set of cycles not in $G$, called the set of \textit{extraneous cycles}. These cycles are directed circuits containing all vertices in $G$, at most $n$ arcs in $G$, and, at least one arc not in $G$, perhaps all arcs not in $G$.\

\textit{About cycles and permutations:} A cycle is a permutation of vertex labels that starts and ends at vertex $n+1$. Cycles and extraneous cycles are coded as permutations and extraneous permutations, respectively.\

\textit{About mapping cycles to permutations:} Assign $p_{u,i}$=1 if and only if the $i^{th}$ arc in a cycle or extraneous cycle of $G$ enters vertex $u$, starting and ending at vertex $n+1$..\ 

\begin{Defn}\label{IND}An \textit{inducer} is the set $\{p_{u,i},p_{v,j}\}$, and is notation for `the $i^{th}$ arc in a cycle or extraneous cycle enters vertex $u$ and the $j^{th}$ arc in a cycle or extraneous cycle enters vertex $v$', synchronizing two sequence positions of a cycle with two vertices of $G$.\end{Defn}\

An interpretation of a cycle in $G$ is usually in terms of arcs used in a graph that connect vertices and orders the way in which vertices are `visited' i.e. a description of a particular permutation of vertex labels. Observe how $\{p_{u,i},p_{v,i+1}\}$  codes all cycles that explicitly use arc $(u,v)$ interpreted as coding for all cycles whose path length from $u$ to $v$ is one. If we instead were to write inducer  $\{p_{u,i},p_{v,j}\}$ as $\{p_{u,i},p_{v,i+m}\}$ we can more easily see that this is code for the set of $n$-permutations matrices whose corresponding Hamilton cycle path length from $u$ to $v$ is $m$. This is an important note referenced in section \ref{sec:ExclusionSet}.

\begin{Defn}\label{INDD}$Perm(\{p_{u,i},p_{v,j}\})$ is the set of permutations satisfying $p_{u,i}=p_{v,j}$=1 and `$\{p_{u,i},p_{v,j}\}$ \textit{induces} a set of {($n$-2)!} permutations'.\end{Defn}\

A single permutation matrix $P$ is  uniquely defined by a set of $\frac{n(n-1)}{2}$ inducers. i.e. $p\cong\Big\{\{p_{u,i},p_{v,j}\}_l$, $l=1,2,...,\frac{n(n-1)}{2}\Big\}$.\

\begin{Defn}\label{BPP}$\Big\{\{p_{u,i},p_{v,j}\}_l$, $l=1,2,...,\frac{n(n-1)}{2},...,k\Big\}$ is said to be a \textit{cover} of $P$ and $P$ is also said to be covered by a set of inducers.\end{Defn}\

\begin{Defn}\label{PP}$\mathcal{P}$ is the set of all $\frac{n^2(n-1)^2}{2}$ inducers.\end{Defn}\

\begin{Defn}\label{BYPP}$F=\big\{\{p_{u,i},p_{v,j}\}_1,\{p_{u,i},p_{v,j}\}_2,...,\{p_{u,i},p_{v,j}\}_{\mathcal{M}}\big\}\subseteq\mathcal{P}$ is called an \textit{induction} set.\end{Defn}\
 
\begin{Defn}\label{BYPYP}$Perm(F)=Perm( \{p_{u,i},p_{v,j}\}_1)\cup Perm(\{p_{u,i},p_{v,j}\}_2)\cup,...,\cup Perm(\{p_{u,i},p_{v,j}\}_\mathcal{M})$, and $F$ is said to induce all of these permutations.\end{Defn}\

\begin{Defn}\label{EG}Let an instance of $G$ be given. Exclusion set $E$ is an induction set such that $P$ is induced by $E$ if and only if $P$ is in correspondence with an extraneous cycle.\end{Defn}\

An exclusion set can be constructed by noting that an extraneous cycle contains at least one arc not in $G$ e.g. $(u,v)$. The complete set of extraneous permutations corresponding to these extraneous cycles is induced by $\{\{p_{u,k}, p_{v,k+1}\}$, $k=1,2,...,n-1\}$. By indexing $k$, arc $(u,v)$ can play the role of {($n$-1)} sequence positions in disjoint sets of extraneous cycles. Considering all $\mathcal{O}(n^2)$ arcs not in $G$, each playing the role of all $\mathcal{O}(n)$ possible sequence positions, it's possible to construct the set of extraneous permutations corresponding to the set of extraneous cycles coded by the union of $\mathcal{O}(n^3)$ $\{p_{u,k},p_{v,k+1}\}$. We construct set $E\subset\mathcal{P}$ in Algorithm \ref{alg:ExcSet} in section  \ref{sec:ExclusionSet} to be $\mathcal{O}(n^4)$, that accounts for some sets of paths longer than one arc not in $G$.\

\subsection{Closure Set $\bar{E}$, Complement of Closure Set $\bar{E}$ \& Perm($\bar{E}$), and Weak Closure Set $\tilde{E}$}
\label{sec:EbarTheory}\setcounter{equation}{-1}\setcounter{equation}{0}

\subsubsection{Closure Set $\bar{E}$, Uniqueness and Link to Hamiltonicity}
\label{ssec:CLEU}\setcounter{equation}{-1}\setcounter{equation}{0}\
Note that distinct exclusion sets can induce the same set of extraneous permutations e.g. two distinct non-Hamiltonian graphs.\
\begin{Defn}\label{equivsets}Two exclusion sets $E_{_1}$ and $E_{_2}$ are equivalent $E_{_1} \equiv E_{_2}$ $\Leftrightarrow$ Perm($E_{_1}$)=Perm($E_{_2}$).\end{Defn} 
\begin{Defn}\label{eqclass}The equivalence class of $E$ is $\mathcal{E}(E)$, the set of all exclusion sets equivalent to $E$.\end{Defn}

\begin{Defn}\label{IG}A closure (set) of $E$ is $\bar{E} \in \mathcal{E}(E)$ of maximum cardinality.\end{Defn}\

\begin{lemm}\label{clos2}Let $\bar{E}_{_1}$, $\bar{E}_{_2}$  $\in$ $\mathcal{E}(E)$. Then $\bar{E}_{_1} = \bar{E}_{_2}$ .\end{lemm}

A closure set is unique and Lemma \ref{clos2} is easy to understand. Imagine the task of exhaustively enumerating each distinct extraneous permutation in $Perm(E_{_1})$=$Perm(E_{_2})$, listed as a row in a table whose columns are the set of all $\frac{n^2(n-1)^2}{2}$ candidate inducers $\{p_{u,i},p_{v,j}\}\in\mathcal{P}$. For each extraneous permutation, assign the column entry corresponding to candidate inducer $\{p_{u,i},p_{v,j}\}$ to be one if and only if $p_{u,i}=p_{v,j}$=1. Build \textit{the} closure set as follows.  For each column, add inducer $\{p_{u,i},p_{v,j}\}$ to the closure set if the column contains {($n$-2)!} ones, yielding one largest set. It follows that $\bar{E} \in \mathcal{E}(E)$ exists and is unique, and $\bar{E} = \mathcal{P}$ if and only if a graph is non-Hamiltonian i.e the set of cycles not in $G$ is the set of all cycles. This is the converse of Theorem \ref{earth} below. $\bar{E}$ might be viewed as a certificate that validates whether or not $Perm(E) = Perm(\mathcal{P})$, and we note that the Hamilton cycle decision problem is therefore a special case of deciding arbitrary $F\,^?_\equiv \, \mathcal{P}$.\\

\begin{theo}\label{earth}$G$ is Hamiltonian $\Leftrightarrow$ $\bar{E} \subset\mathcal{P}$ $\Leftrightarrow$ $|\bar{E}|\,\,<\frac{n^2(n-1)^2}{2}$.\end{theo}\

\newpage

\subsubsection{Complement of Closure Set $\bar{E}$ and Complement of Perm($\bar{E}$)}
\label{ssec:MTCLEU}\setcounter{equation}{-1}\setcounter{equation}{0}\

\begin{Defn}\label{EC}The complement of $\bar{E}$ is $\bar{E}^{^C}=(\bar{E})^{^C}=\mathcal{P}\backslash\bar{E}$.\end{Defn}

\begin{Defn}\label{EPC}The complement of $Perm(\bar{E})$ is $Perm^{^C}(\bar{E})=(Perm(\bar{E}))^{^C}=Perm(\mathcal{P})\backslash Perm(\bar{E})$.\end{Defn}\\

$Perm^{^C}(\bar{E})$ and $\bar{E}^{^C}$ are related. Let $p\in Perm^{^C}(\bar{E})$, $p\cong\Big\{\{p_{u,i},p_{v,j}\}_l$, $l=1,2,...,\frac{n(n-1)}{2}\Big\}$. Each $\{p_{u,i},p_{v,j}\}_l\in\bar{E}^{^C}$, otherwise there exists $\{p_{u,i},p_{v,j}\}_l\in\bar{E}$ $\Rightarrow$ $p\in Perm(\bar{E})$, a contradiction. Therefore if $p\in Perm^{^C}(\bar{E})$ then $P$ is covered by a set of inducers in $\bar{E}^{^C}$. But what about the converse? Does there exist an inducer in $\bar{E}^{^C}$ that is not part of a covering of any $p \in Perm^{^C}(\bar{E})$? Suppose that $\{p_{u,i},p_{v,j}\}\in\bar{E}^{^C}$ is such an inducer. Then all $P$ satisfying $\{p_{u,i},p_{v,j}\}=1$ are members of Perm($\bar{E}$) $\Rightarrow$  $\{p_{u,i},p_{v,j}\}\in\bar{E}$ by the definition of closure, a contradiction. This leads to Lemmae \ref{coverlemm1} and \ref{coverlemm2}.\  

\begin{lemm}\label{coverlemm1}$G$ is Hamiltonian $\Leftrightarrow$ $Perm^{^C}(\bar{E}) \ne \emptyset $ $\Leftrightarrow$ $\bar{E}^{^C} \ne \emptyset.$ \end{lemm}
\begin{lemm}\label{coverlemm2}Let $p\cong\Big\{\{p_{u,i},p_{v,j}\}_1,\{p_{u,i},p_{v,j}\}_l, l=2,...,\frac{n(n-1)}{2}\Big\}$. Thus $p\in Perm^{^C}(\bar{E})$ $\Leftrightarrow$

$\{p_{u,i},p_{v,j}\}_1\in\bar{E}^{^C}$ $\Leftrightarrow$ $\Big\{\{p_{u,i},p_{v,j}\}_1,\{p_{u,i},p_{v,j}\}_l, l=2,...,\frac{n(n-1)}{2}\Big\}\subseteq \bar{E}^{^C}.$\end{lemm}\

Simple consequences of Lemmae \ref{coverlemm1} and \ref{coverlemm2} are: 1) $G$ is non-Hamiltonian if and only if $\bar{E}^{^C}=\emptyset$\, 2) $G$ has precisely one cycle if and only if $|\bar{E}^{^C}|=\frac{n(n-1)}{2}$, 3) $G$ has more than one cycle if and only if $|\bar{E}^{^C}|>\frac{n(n-1)}{2}$.\

\subsubsection{Weak Closure Set $\tilde{E}$ and Prelude to WCA}
\label{ssec:PR}\setcounter{equation}{-1}\setcounter{equation}{0}\
We introduce the concept of weak closure set $\tilde{E}$, an approximation of $\bar{E}$. Recall that $\tilde{E}$ was temporarily defined in section \ref{sec:IntroAlgo}.\

\begin{Defn}\label{WG}Let an instance of $G$ be given, together with $E$. A weak closure (set) of $E$ is $\tilde{E} \in \mathcal{E}(E)$, where $E \subseteq \tilde{E} \subseteq \bar{E}$.\end{Defn}\

\begin{theo}\label{earth1}If $\tilde{E}=\mathcal{P} \Leftrightarrow |\tilde{E}|=\frac{n^2(n-1)^2}{2}$ then $G$ is non-Hamiltonian.\end{theo}\ 

\section{Construction of Exclusion Set $E$}
\label{sec:ExclusionSet}\setcounter{equation}{-1}\setcounter{equation}{0}\
See Algorithm \ref{alg:ExcSet} below. Recall that $G$ is connected and that it's sufficient for an exclusion set to account only for arcs not in $G$ i.e. recall the comments following presentation of System 1. Also recall that inducer $\{p_{u,i},p_{v,j}\}$ ($i<j$) codes sets of permutations corresponding with cycles whose path length from $u$ to $v$ is $j-i$. Thus for arc $(u,v)$ not in $G$, elements of $E$ are inducers of the form $\{p_{u,i},p_{v,i+1}\}$, and we also include additional inducers of the form $\{p_{u,i},p_{v,i+m}\}$ whenever it's possible to account for no paths of length $m$ in $G$, from vertex $u$ to vertex $v$. We do this by implementing Dijkstras algorithm with equally weighted arcs to find minimal length paths between all pairs of vertices, coded to return $m=n+1$ if no path exists. We account for all paths of length one not in $G$ (arcs not in $G$), and, all paths of length two not in $G$ by temporarily deleting the arc between adjacent vertices.\

Begin as follows. If $u$ is adjacent to $v$ then temporarily delete arc ($u$,$v$) and apply Dijkstras algorithm to discover a minimal path of length $m>1$. No paths of length $k$ can exist, $k=1,...,m-1$ and inducers are discovered that 1) for $k=1$ and $u$ not adjacent to $v$ correspond with arcs in extraneous cycles\footnotemark[5], and 2) for $k>1$ corresponding with paths of length $k$ in extraneous cycles. Thus while coding for all arcs not in $G$ is sufficient to exclude integer solutions of System 2 in the case of non-Hamiltonian $G$, we also code for path segments not in $G$ in order to bolster our input set of $q$ variables to be set zero.\
\footnotetext[5]{Using projection and lifting techniques, based upon ideas presented here, a compact formulation whose image polytope is the convex hull of extraneous permutations of $G$ was first constructed in \cite{gs04} and developed further in \cite{gismo13a}.} 

The general case is described in section \ref{ssec:TE}. But two special cases arise. \textit{Case 1. Last arc in cycle}: Recall that every $n+1^{th}$ arc in a cycle enters vertex $n+1$ by definition. Therefore observe arcs ($u$,$n+1$) not in $G$, temporarily deleted or otherwise, noting how corresponding sets of extraneous cycles can be coded by permutations for which the $n^{th}$ arc in a cycle enters vertex $u$ i.e. $p_{u,n}=1$. This is the case for $k$=1, and $u$ not adjacent to $v$ when Dijkstras algorithm returns $m=2$. If Dijkstras algorithm returns $m=3$, then again for $k$=1 and if $u$ is not adjacent to $v$ code $p_{u,n}=1$, and for $k$=2, no paths of length two exist and these sets of extraneous cycles can be coded by permutations for which the $n-1^{th}$ arc in a cycle enters vertex $u$ i.e. $p_{u,n-1}=1$. Continuing in this way, code all possible $n+1-k^{th}$ arcs in extraneous cycles, in paths of length $k$ not in $G$, to enter vertex $u$ i.e. $p_{u,n+1-k}=1$, $k=1,2,...,m-1$. \textit{Case 2. First arc in cycle}: Recall every first arc in every cycle exits vertex $n+1$. Observe and code all arcs ($n+1$,$v$) in extraneous cycles, in paths of length $k$ not in $G$ by coding all possible $k^{th}$ arcs to enter vertex $v$ i.e. $p_{v,k}$=1, $k=1,2,...,m-1$.\

Recall that $G$ is strongly connected. But if an arc is temporarily deleted, it's possible for no path to exist between a given pair of vertices. This useful information indicates an arc is essential under the assumption of the existence of a Hamilton cycle. In case 1, this implies that a particular $p_{u,n}$ is necessary, and by integrality must be at unit level in every assignment of variables, assuming the graph is Hamiltonian (until deduced otherwise, if ever). Thus all other $P$'s in the same row (and column) can be set at zero level. This is accounted in the algorithm. Recall that $m=n+1$ in the case that Dijkstras algorithm returns no minimal path. The $k$ loop appends the necessary set of $\{p_{v,j},p_{u,n+1-k}\}$ inducers to $E$ effectively setting variables in blocks ($u$,1,) through ($u$,$n-1$) to zero. When the WCA is implemented, $p_{u,n}$ must attain unit level via double stochastity, and this implies that the other $P$'s in the same column are deduced to be at zero level, again via double stochastity. Similarily for Case 2. In the general case, it's also possible for no path to exist between a given pair of vertices ($u$,$v$) (when an arc is temporarily deleted). Under the assumption of the existence of a Hamilton cycle, this arc is essential and can play the role of sequence position 2 through $n$-1 and so in each case, all complementary row (and column) inducers are assigned to $E$. When the WCA is implemented, a single $q$ variable remains in each row and therefore is  equated with that block's $p_{u,i}$ variable via `scaled' double stochastity \textit{within} the block i.e. rows and columns in the block sum to $p_{u,i}$. All complementary $q$ variables in the corresponding column are therefore set 0 in each block. Refer to section \ref{sec:MaxCoverAlgo}, noting how inducers correspond with $q$ variables constrained so that each row and each column sums to that block's $p$ variable i.e.  scaled double stochastity in sub-blocks of a larger doubly stochastic $\mathbb{Q}$ matrix. See Figure \ref{Qmat} and explore how inducers relate to one another via double stochastity and scaled double stochastity. In summary, essential arcs also contribute to new information by adding their complementary row / column inducers to $E$.\

\begin{algorithm}
\KwIn{Arc Adjacency matrix for $G$.}
\KwOut{$E$.}
$E\leftarrow\emptyset$\;
Case 1: \For{$u=1,2,...,n$}{
$Arc \leftarrow G(u,n+1); G(u,n+1) \leftarrow 0$; $m \leftarrow$ DijkstrasAlgorithm($G$,$u$,$n+1$)\;
\For{$k=Arc+1,Arc+2,...,m-1$}{
$E\leftarrow E\cup \{p_{v,j},p_{u,n+1-k}\}, v=1,2,...,n, v\neq u; j=1,2,...,n, j\neq n+1-k$\;
Note: \textit{It's sufficient to code only for $j<n+1-k$)};
}
$G(u,n+1) \leftarrow Arc$\;
}
Case 2: \For{$v=1,2,...,n$}{
$Arc \leftarrow G(n+1,v); G(n+1,v) \leftarrow 0$; $m \leftarrow$ DijkstrasAlgorithm($G$,$n+1$,$v$)\;
\For{$k=Arc+1,Arc+2,...,m-1$}{
$E\leftarrow E\cup \{p_{v,k},p_{u,i}\}, u=1,2,...,n, u\neq v; i=1,2,...,n, i\neq k$\;
Note: \textit{It's sufficient to code only for $k<i$)};
}
$G(n+1,v) \leftarrow Arc$\;
}
General Case. \For{$u=1,2,...,n$}{
\For{$v=1,2,...,n; v\neq u$}{
$Arc \leftarrow G(u,v); G(u,v) \leftarrow 0$; $m \leftarrow$ DijkstrasAlgorithm($G$,$u$,$v$)\;
\For{$k=Arc+1,Arc+2,...,m-1$}{
$E\leftarrow E\cup \{p_{u,l},p_{v,l+k}\}, l=1,2,...,n-k$\;
}
$G(u,v) \leftarrow Arc$\;
}
}
RETURN $E$\;
\
\caption{Construction of Exclusion Set $E$}
\label{alg:ExcSet}
\end{algorithm}
\clearpage
\section{The Weak Closure Algorithm (WCA)}
\label{sec:MaxCoverAlgo}\setcounter{equation}{-1}\setcounter{equation}{0}\
The algorithm deduces values of variables of System 2 by contradiction. Variables are tested at zero and/or unit levels and if System 2 is infeasible, we deduce the value of the variable as the opposite of its assigned test value. This is Algorithm \ref{alg:SteveWCA} below, supported by Routines 1 through 4 (Appendix).\

In line 1, we test System 2 for infeasibility (non-Hamiltonicity), setting $q$ variables to zero / unit level via \textbf{ImplementClosure} (Routine 1). We first test for no match with respect to the set of $p$ variables, in which case we return the decision `Infeasible'. Otherwise we exhaustively minimize / maximize each $\tilde{q}_{u,i,v,j}$ (in correspondence with each $\{\tilde{p}_{u,i},\tilde{p}_{v,j}\} \in \mathcal{P}\backslash \tilde{E} \backslash \tilde{F}$) in order to decide whether or not $\tilde{q}_{u,i,v,j}$ can attain zero and/or unit value in a feasible solution of System 2 (replace $E$ with $\tilde{E}$). If it cannot attain unit level (zero level) then $\tilde{q}_{u,i,v,j}$ is set to zero level (unit level), and $\{\tilde{p}_{u,i},\tilde{p}_{v,j}\}$ is added to $\tilde{E}$ ($\tilde{F}$). Note that an assignment to unit level invokes many more $\tilde{q}$ variable zero assignments. See \textbf{Assign\_$\tilde{q}_{u,i,v,j}=1$} (Routine 4). We repeat this process until all $q$ variables can attain both zero and unit level in feasible solutions of System 2. However, if Routine 1 returns the decision `Infeasible', the WCA deduces non-hamiltonicity. If Routine 1 returns the decision `Feasible Integer Solution', the WCA deduces Hamiltonicity. Otherwise Routine 1 returns `Feasible Fractional Solution' and the WCA continues to a nested version of what we just described.\

In this nested stage, we systematically assign each $\tilde{q}_{u,i,v,j}$ to be 0 and/or 1 (in correspondence with each $\{\tilde{p}_{u,i},p_{v,j}\}  \in \mathcal{P}\backslash \tilde{E} \backslash \tilde{F}$). Recall that a feasible solution with this assignment is guaranteed to exist since this is the exit criteria for Routine 1. We then assume the existence of a feasible integer solution and test our hypothesis. Thus we test all $\tilde{\tilde{q}}_{u,i,v,j}$ re: zero and unit level just as we did in Routine 1, we code this as \textbf{TestAssumption} (Routine 3). This is the very same closure technique as implemented in Routine 1. We then exhaustively test and assign each of the remaining variables at zero / unit levels. If Routine 3 returns the decision `Infeasible', the WCA deduces the correct assignment of $\tilde{q}_{u,i,v,j}$ as the opposite of its assumed assignment, and the WCA starts again with this new information. See the code that follows lines 2 \& 3. If Routine 3 returns the decision 'Feasible Integer Solution', the WCA deduces Hamiltonicity. Otherwise Routine 3 returns `Feasible Fractional Solution' and we make no deductions about $\tilde{q}_{u,i,v,j}$ and we test another $\tilde{q}_{u,i,v,j}$ until all variables have been tested.\

When the WCA sets a $q$ variable to unit level, the set of complementary row and column $q$ variables are set to zero for computational convenience (size and speed) in Routines 2 \& 4. There's no need to test them.  For non-Hamiltonian $G$ and knowing that there exist no 0/1 solutions, we hope to speed-up the WCA.\

Finally, the WCA can exit `Undecided' with a set of $\tilde{q}_{u,i,v,j}$ that attain both zero and unit levels in feasibile fractional solutions of System 2 (replace $E$ with $\tilde{E}$). For non-Hamiltonian $G$, $\tilde{E} \subset \bar{E}=\mathcal{P}$ even though $\tilde{E}\equiv\mathcal{P}$. Thus $\tilde{E}^{^C}=\mathcal{P}\backslash \tilde{E}$ covers no $P$, the basis for some of the ideas presented in section \ref{sec:TermFeasWCA}.\

\clearpage
\begin{algorithm}[h!]
\SetKwBlock{Block}{begin}{end}
\KwIn{(System 2, $E$, Matrix $Q$)} 
\KwOut{(Decision)}
Matrix $\tilde{Q} \leftarrow $ Matrix $Q \,\, \cup \,\, \{q_{u,i,v,j}\leftarrow 0$, $\textbf{foreach }\{p_{u,i},p_{v,j}\} \in E\}$; $\tilde{LP}\leftarrow$ System 2; $\tilde{E} \leftarrow E$; $\tilde{F} \leftarrow \emptyset $\;

\While{(not true that lines 1, 2 or 3 return `Feasible Integer Solution')}{
\textbf{Start Again}\;
\nl \textbf{ImplementClosure}($\tilde{LP}$,$\tilde{E}$,$\tilde{Q}$,$\tilde{F}$)\;
\eIf{Decision = `Infeasible'}
{EXIT non-Hamiltonian}
{\ForEach{$\{\tilde{p}_{u,i},\tilde{p}_{v,j}\}\in \mathcal{P}\backslash\tilde{E}\backslash\tilde{F}$}{
        $\tilde{\tilde{Q}} \leftarrow \tilde{Q}$; $\tilde{\tilde{LP}}\leftarrow\tilde{LP}$; $\tilde{\tilde{E}} \leftarrow \tilde{E}$; $\tilde{\tilde{F}} \leftarrow \tilde{F}$\;
\textbf{Assume\_$\tilde{q}_{u,i,v,j}=1$}($u,i,v,j$,$\tilde{\tilde{LP}}$, $\tilde{\tilde{E}}$, $\tilde{\tilde{Q}}$, $\tilde{\tilde{F}}$)\;
\nl \textbf{TestAssumption}($\tilde{\tilde{LP}}$,$\tilde{\tilde{E}}$,$\tilde{\tilde{Q}}$,$\tilde{\tilde{F}}$)\;
         \eIf{Decision = `Infeasible'}
         {\textbf{Assign\_$\tilde{q}_{u,i,v,j}=0\equiv$}  $\{\tilde{LP}\leftarrow\tilde{LP}$ \textit{and} $\{\tilde{Q}\leftarrow\tilde{q}_{u,i,v,j}\leftarrow 0\}$; $\tilde{E}\leftarrow\tilde{E}\cup\{\tilde{p}_{u,i},\tilde{p}_{v,j}\}\}$\;
           \textbf{goto} Start Again\;}
        {
           $\tilde{\tilde{Q}} \leftarrow \tilde{Q}$; $\tilde{\tilde{LP}}\leftarrow\tilde{LP}$; $\tilde{\tilde{E}} \leftarrow \tilde{E}$; $\tilde{\tilde{F}} \leftarrow \tilde{F}$\;

\textbf{Assume\_$\tilde{q}_{u,i,v,j}=0\equiv$} $\{\tilde{\tilde{LP}}\leftarrow\tilde{LP}$ \textit{and} $\{\tilde{\tilde{Q}}\leftarrow\tilde{\tilde{q}}_{u,i,v,j}\leftarrow 0\}$; $\tilde{\tilde{E}} \leftarrow \tilde{E}\cup\{\tilde{\tilde{p}}_{u,i},\tilde{\tilde{p}}_{v,j}\}\}$\;
        
        \nl \textbf{TestAssumption}($\tilde{\tilde{LP}}$,$\tilde{\tilde{E}}$,$\tilde{\tilde{Q}}$,$\tilde{\tilde{F}}$)\;
         \eIf{Decision = `Infeasible'}
         {\textbf{Assign\_$\tilde{q}_{u,i,v,j}=1$}($u,i,v,j$,$\tilde{LP}$, $\tilde{E}$, $\tilde{Q}, \tilde{F}$)\;
           \textbf{goto} Start Again\;}
{$\tilde{q}_{u,i,v,j}$ attains 0 and 1 in Feasible Fractional Solutions of $\tilde{LP}$.\;
}
}
}
All $\tilde{q}_{u,i,v,j}$ attain 0 and 1 in Feasible Fractional Solutions of $\tilde{LP}$.\;     
EXIT Undecided   
         }
         }
{EXIT Hamiltonian}
\
\caption{The Weak Closure Algorithm}
\label{alg:SteveWCA} 
\end{algorithm}

\section{Results and Discussion}
\label{sec:ResultsDisc}\setcounter{equation}{-1}\setcounter{equation}{0}\
Over 100 non-Hamiltonian graphs (10 through 104 vertices) and 2000 randomized 31 vertex non-Hamiltonian graphs are tested and correctly decided non-Hamiltonian via the WCA, coded in F77, run on a Mac Mini Desktop using Absoft Fortran 13.01. A summary set of details for 20 of these graphs are shown in table \ref{fourdubs} below\footnotemark[2]. The heading \# $p_{u,i} (q_{u,i,v,j})$ refers to the count of non-zero components in $Q$ after implementing Algorithm ref{alg:ExcSets}, before implementing the WCA. Note that $p_{u,i}=q_{u,i,u,i}$, and we only count $q_{u,i,v,j}$ $i<j$. The heading $|\tilde{E}^{^C}|\le$ refers to an upper bound on  $|\tilde{E}^{^C}|$ for 11 selected graphs, each modified to include the tour $1-2-...-n-n+1$, simply to observe $\tilde{E}^{^C}$. Two of these graphs are also hypohamiltonian, and the count in parentheses is an upper bound on $|open(V)|$ after removing a vertex.

\begin{table}[h!] 

\caption{Non-Hamiltonian Graph Test Runs of the WCA} 
\label{fourdubs}
{\centering     

\begin{tabular}{lccc}  
Name of Graph & \# Vertices (Edges) & \# $p_{u,i} (q_{u,i,v,j})$ & $|\tilde{E}^{^C}| \le$  \\ [1ex]    
\hline                
Petersen Snark & 10 (15) & 57 (858) &  722\\  
Herschel Graph & 11 (18) & No 2-Factor &  1,980\\
A Kleetope & 14 (36) & 147 (8,166) &  5,809\\  
Matteo\cite{Leder16Other} & 20 (30) & 275 (26,148) &  27,093\\
Coxeter & 28 (42) & 597 (136,599) & 135,453(1,241)$^1$\\  
Graph \#3337 Snark\cite{bcgm2013} & 34 (51) & 897 (308,234) & 335,697\\
Zamfirescu Snark & 36 (54) & 983 (363,987) & 7,749\\
Barnette-Bosák-Lederberg & 38 (57) & 1077 (440,318) & 96,834\\
A Hypohamiltonian & 45 (70) & 1,656 (1,109,738) & 296,668 (29,724)$^1$\\
Tutte & 46 (69) & 1,649 (1,060,064) & 436,250\\
A Grinberg Graph & 46 (69) & 1,737 (1,204,722) & - Not yet run -\\  
Georges & 50 (75) & 2037 (1,701,428?) & - Not yet run -\\
Szekeres Snark & 50 (75) & 2045 (1,718,336) & - Not yet run -\\
Watkins Snark & 50 (75) & 2051(1,708,987) & - Not yet run -\\
Ellingham-Horton & 54 (81) & 2,315 (2,135,948)  & 1,045,041\\
Thomassen & 60 (99) & 3,105 (4,071,600) & - Not yet run -\\  
Meredith & 70 (140) & 4,221 (7,526,996) & - Not yet run -\\
A Flower Snark & 76 (114) & 4,851 (9,720,420) & - Not yet run -\\  
Horton & 96 (144) & 8,205 (29,057,118) & - Not yet run -\\  
A Goldberg Snark &104 (156) & 9,339 (37,802,124) & - Not yet run -\\  
[1ex]    
\hline   
\end{tabular} 

{\footnotesize $^1$ Hypohamiltonian. We also confirmed existence of non-empty $\tilde{E}^{^C}$ after removing a vertex and re-running the WCA.}\

}\
\end{table}

\footnotetext[2]{In addition the those graphs listed in Table \ref{fourdubs}, the WCA closed 10 more varieties of Goldberg, Flower, Blanusa, Loupekine and Clemins-Swart Snarks, and over 90 more House of Graphs graphs ranging in vertex count from 26 through 38 vertices.}

\subsection{About Tweaking Input to the WCA}
\label{sec:TweWCA}\setcounter{equation}{-1}\setcounter{equation}{0}Plenty of special case graphs exist for which some pre-processing may help improve run-times. For example, if $G$ is undirected and Hamiltonian, $G$ can be made directed by deleting an arc in one direction, and remains Hamiltonian. Some arcs may be better candidates for deletion than others re: performance. Assuming that the WCA can't always close $E$ for non-Hamiltonian $G$ (otherwise \textbf{P=NP}), a lucky choice of arc might also cause $E$ to close where it did not originally close.\

For three regular undirected Hamiltonian graphs, we can choose a vertex and delete an edge (two arcs in opposite directions). Two degree two vertices result. If $G$ remains Hamiltonian, we can again impose direction through one of these vertices, causing six arcs to be deleted (improved performance) and this directed graph remains Hamiltonian. Repeating this process for each of three edges, it must be the case that $E$ cannot close for at least one of these cases, otherwise $G$ is non-Hamiltonian. Thus for non-Hamiltonian $G$, we can hope that for the right choice of vertex, in all three cases $E$ becomes closed i.e. deciding non-Hamiltonicity. For each of these three cases, we can also impose direction in two different ways, on the second degree two vertex. This causes up to six more arcs to be deleted for each way. For non-Hamiltonian $G$, we hope that $E$ becomes closed for each way, for each of three cases, for a lucky choice of vertex. Note that we performed and validated these tests and ideas for the Petersen graph (non-Hamiltonian), and the Petersen graph with an extra edge (Hamiltonian).\

\subsection{About Generalizing the WCA}
\label{sec:GenWCA}\setcounter{equation}{-1}\setcounter{equation}{0}As currently implemented, the WCA tests all $\{p_{u,i},p_{v,j}\}$ to be at 0/1 level. In general, $p$ variables can be triples, quadruples and so on. A generalized WCA involves testing all generalized $p$ variables e.g. $\{p_{u,i},p_{v,j},p_{w,k}\}$ or $\{p_{u,i},p_{v,j},p_{w,k},p_{x,l}\}$ to be at 0/1 level etc., invoking higher order models until there are as many variables as permutations. Low order implementations might be useful for some classes of non-Hamiltonian graphs, assuming there exist useful complexity bounds regarding the amount of computation sufficient to \textit{guarantee} collapse of corresponding polyhedra. We note that higher order models capture increasingly more intricate and specialized information. For example, if we deduce that $\{p_{u,i},p_{v,j},p_{w,k}\}=0$, we interpret that there are no cycles for which the $i^{th}$ arc enters vertex $u$ and the $j^{th}$ arc enters vertex $v$ and the $k^{th}$ arc enters vertex $w$. This suggests an approach toward building tough counter examples for the WCA i.e. construct a graph that is dense with little pieces of paths whose corresponding variables are not linearly dependent. To be clear, in the \textit{General Case} of Algorithm \ref{alg:ExcSet}, fewer variables are set zero if we detect shortest paths of low length. While the structure of the graph decides how equations relate variables to one another, at face value, this might be a good working rule.\

\newpage

\subsubsection{How to Create an Exclusion Set That Models the Graph non-Isomorphism Decision Problem}
\label{sec:coNPmods}\setcounter{equation}{-1}\setcounter{equation}{0} From \cite{gismo13a}, `\textit{These [modelling] techniques might be applicable to the graph isomorphism [decision] problem. Recall that graph $G$ is isomorphic to graph $H$ if and only if there exists permutation matrix $P$ such that $P^TGP=H$ ...  to investigate the possibility of modelling the set of permutations satisfying either $P^TGP = H$ or $P^TGP \ne H$.}'. We can create $E$ as follows. Construct $P^TGP=H$ i.e. the set of $n^2$ equations composed of sums of products of pairs of components of $P$, their sum being either 0 or 1, depending on $H$. Each term of the form $p_{u,i}\cdot p_{v,j}$ in each of the equations set zero is non-negative and must therefore be at zero level in every solution of $P^TGP = H$. Define $E$ as the set of  corresponding $\{p_{u,i},p_{v,j}\}$. It must be the case that for each $\{p_{u,i},p_{v,j}\}$, either $p_{u,i}=0$ or $p_{v,j}=0$ or $p_{u,i}=p_{v,j}=0$ in System 2 i.e. assign $q_{u,i,v,j}=p_{u,i}\cdot p_{v,j}$. Recall that by assigning corresponding $q_{u,i,v,j}=0$ in System 2, precisely those $P$ solutions of System 2 satisfying $q_{u,i,v,j}=p_{u,i}\cdot p_{v,j}=p_{u,i}=p_{v,j}=1$ are excluded. The complement set of $P$ solutions of System 2 is therefore the set of all possible $P$ that map precisely all zeroes of $G$ to all zeroes of $H$. But if $P$ exists, $P^TGP$ must also preserve the arc count of $G$, by definition and it therefore follows that, if $P$ is a solution of System 2 then $P^TGP=H$ i.e. if $P^TGP \ne H$, there exists no $P$ solution of System 2. Thus $E$ is  sufficient to exclude all $P$ solutions to System 2. These ideas follow directly from study of a proposed model of the graph non-isomorphism problem. That is, the set of extrema of a compact formulation of the union of sets of Birkhoff polyhedra, each polytope the solution set of assignment constraints in correspondence with each assignment of $p_{u,i}=p_{v,j}=1$ (using projection and lifting techniques) are not solutions of Systems 2.\

\subsection{About the WCA Terminating With a Feasible LP for non-Hamiltonian $G$}
\label{sec:TermFeasWCA}\setcounter{equation}{-1}\setcounter{equation}{0}\
In these cases, the LP can be made infeasible by choosing additional $q_{u,i,v,j}$ variables to be set zero where the goal of the WCA is now proposed as a tool to verify non-Hamiltonicity. For example, choose a row/column of matrix $Q$ for which many of the $q_{u,i,v,j}$ variables have already been set zero and begin to hopefully `set in motion' a remaining cascade of additional $q_{u,i,v,j}$ variables that become zero, via application of the WCA. There may exist a `best way' to do this no matter that the problem remains to confirm that the set of additional $q_{u,i,v,j}$ variables that we choose to set zero in fact can be verified zero via some polynomial time means - assuming the WCA is to be useful. In this way, the WCA can be used as a tool to provide hints about additional $q_{u,i,v,j}$ variables to be investigated and `a priori' set zero. That is, while not originally included in $E$, if discovered, they can now be included and application of the WCA terminates infeasible. This revised set of input constitutes `a polynomial sized set of correctly guessed information' necessary to verify a correctly guessed non-Hamiltonion $G$ in polynomial time. This amounts to studying a \textbf{coNP}-complete problem in $\mathcal{O}(n^4)$ variables as a set of smaller instances of `core' (relative to the WCA) \textbf{coNP}-complete problems in the case of verifying a correctly guessed set of variables to be set zero.\

More generally, knowing that $E \equiv \tilde{E} \equiv \bar{E}=\mathcal{P}$, it follows that $Perm(\tilde{E})$ is the set of all $P$. Therefore $Perm^C(\tilde{E})=\emptyset$ and by application of the converse of Lemma \ref{coverlemm1},  $\tilde{E}^C \equiv \emptyset$ and therefore covers no $P$. A general approach toward resolving the existence of a cover of $P$ is to assume it exists and show a contradiction, like the WCA, however possible. Maybe after application of the WCA, the problem is reduced in a way that can yield other insights. One thought is to model the new problem as a smaller instance of non-Hamiltonicity and re-apply the WCA. But this `recursive' approach might simply devolve into an exhaustive technique.\ 

We now comment that $C7-21$  \cite{BondMurt} is a non-Hamiltonian graph for which, after application of the WCA we discover that $E$=$\tilde{E}$ and the WCA terminates with a feasible LP i.e. no matter that we know $E\equiv \mathcal{P}$, the WCA can deduce no additional $q$ variables to be set zero. Note that $C7-21$ is an undirected 21 vertex graph. It's composed of a 14 vertex complete graph, and seven wheels. Each of vertices 15 through 21 is the centre of a wheel adjacent to vertices one through seven. The minimum vertex degree is seven.\

\subsection{Next Steps}
\label{sec:PracticalWCA}\setcounter{equation}{-1}\setcounter{equation}{0}\
Three do-able project ideas now follow. \textit{First Project:} It would be interesting to run a sequence of snarks in search of a counter example to help understand where the WCA can fail. Otherwise, if the WCA continues to close $E$, it's equally interesting to model and estimate the order of run-time for detecting non-Hamiltonicity as a function of the number of vertices. \textit{Second Project:} It would be interesting to run a sequence of degree three regular Hamiltonian graphs, in search of the order of run-time as a function of the number of vertices. That is, the WCA cannot close $E$ in these cases and we can model the order of computation by noting run times at the point where no more changes were made to $E$. \textit{Third Project:} Depending on how well the WCA performs in the first two projects, we could create sets of degree-regular randomly Hamiltonian and non-Hamiltonian graphs in order to simply test how well the WCA performs as a function of degree regularity and number of vertices. We would need to also implement a search for a cycle within the cover set in cases where $E$ does not close in order to know if the WCA makes a correct decision.\

\subsubsection{State Matrices}
\label{sec:PracticalWCA4}\setcounter{equation}{-1}\setcounter{equation}{0}\
Perhaps a more general way to study the WCA, applicable to all of the ideas that we present is to view the $Q$ matrix in Figure \ref{Qmat} as a state matrix. The problem then is to create and study the WCA, including higher order implementations of the WCA, as generating a sequence of states. Properties of these state sequences may be predictive of certain kinds of outcomes e.g. whether or not a final state of a zeroed $Q$ matrix can be expected etc. This depends upon the complexity of generating new states that can be recognized as deciding the problem. The idea of investigating recurrence relations between states might also be useful etc., the goal being to recognize / test each state re: deciding Hamiltonicity.\

\subsubsection{Upper Bounds}
\label{sec:PracticalWCA1}\setcounter{equation}{-1}\setcounter{equation}{0}\
Understanding $|E|$ relative to both the WCA (closure) and classes of non-Hamiltonian $G$ is important. Can we associate classes of non-Hamiltonian graphs, $|E|$ and a guarantee of closure? Can we associate classes of non-Hamiltonian graphs with the complexity of computing additional components of $E$ necessary to guarantee closure? How should we study the relationship between $E$, non-Hamiltonian $G$ and the WCA? We propose a structured set of tests of the WCA using large randomized classes of non-Hamiltonian graphs with the intent of establishing some highly probable upper bounds on $|E|$ and a guarantee of closure.\

\subsubsection{Efficiency}
\label{sec:PracticalWCA2}\setcounter{equation}{-1}\setcounter{equation}{0}\
It would be important to investigate other techniques that decide non-Hamiltonicity and compare them to the WCA as currently implemented and also higher order implementations. We can begin these comparisons by studying patterns of successes and failures i.e. the hope is to gain insight about a theory to develop a more robust WCA-like technique. This could involve studies of 1) counter-examples, or 2) three regular graphs (counter examples and successes), or 3) subsets of graphs e.g. snarks etc. We propose the construction of a large set of non-Hamiltonian $G$ for some class of graphs e.g. snarks, and compute, tabulate and organize in some way, minimal $|E|$ sufficient for the WCA to close $E$ for comparison to other methods that decide non-Hamiltonicity. It may be necessary to create computer code that dynamically generalizes the WCA via increased nesting of assignment polyhedra (leading to higher order $|E|$) as discussed above.\

\subsubsection{The WCA in Another Context}
\label{sec:PracticalWCA3}\setcounter{equation}{-1}\setcounter{equation}{0}\
Separate from application, regarding improved understanding of the WCA, it might be useful to investigate the creation of a relaxed compact linear programming formulation of the Hamilton cycle decision problem using $p_{i,j,k}$ variables and triply stochastic matrices defined in a suitable way, where $k$ is the sequence position of arc $(i,j)$ in a cycle. This is a more natural way to understand cycles using permutations. We can again code information about missing arcs and paths from its \textbf{coNP} counter model and create a corresponding WCA.  Maybe there exists an insightful relationship to three regular graphs, another approach toward study of \textbf{NP}-completeness.\

\subsection{Closing Remarks}
\label{sec:PracticalClose}\setcounter{equation}{-1}\setcounter{equation}{0}\
There is one final thought to be appreciated that captures the large view of our intent to study \textbf{coNP}. There exists an underlying recursive relationship between the $n^2-2n+1$ dimensional Birkhoff polytope and its sub-polytope, the $n^2-3n+2$ dimensional TSP polytope. The set of extrema of the TSP polytope is the set of Hamilton cycles of a complete $n$ vertex graph, and is a sub-polytope of the Birkhoff polytope whose set of extrema is the set of all $n$-permutations. But these permutations, via the ideas presented in this paper are in correspondence with the set of all Hamilton cycles of a complete $n+1$ vertex graph, whose convex hull after mapping back to cycles as presented at the outset, is the set of extrema of the TSP polytope of a complete graph on $n+1$ vertices, a sub-polytope of the Birkhoff polytope whose set of extrema is the set of all ($n+1$)-permutations. But these permutations (via the ideas presented in this paper) are in correspondence with the set of all Hamilton cycles of a complete $n+2$ vertex graph, whose convex hull after mapping back to cycles is the set of extrema of the TSP polytope of a complete graph on $n+2$ vertices, a sub-polytope of the Birkhoff polytope whose set of extrema is the set of all ($n+2$)-permutations, and so on. If we imagine that the Birkhoff polytope is \textit{the example} of symmetry and tractability while the TSP polytope is \textit{the example} of asymmetry and intractability, we wonder if there might exist any useful insights regarding properties of the TSP polytope for a graph on $n+1$ vertices that can be discovered by studying the Birkhoff polytope that models $n$-permutations i.e. a dynamic matching algorithm and the $Q$ matrix?\

\clearpage
\section{Appendix}
\label{sec:Appendif}\setcounter{equation}{-1}\setcounter{equation}{0}\
\begin{algorithm}[h!]
\SetKwBlock{Block}{begin}{end}
\KwIn{($\tilde{LP}$, $\tilde{E}$, $\tilde{Q}$, $\tilde{F}$)} 
\KwOut{(Decision, $\tilde{LP}$, $\tilde{E}$, $\tilde{Q}$, $\tilde{F}$)}
\While{(not true that lines 1 or 2 return `Infeasible' or `Feasible Integer Solution')}{
\textbf{Top\_p}\;
       \For{$k,l=1,2,...,n; $}{
            $P_{k,l}$ $\leftarrow$ 1\;
            \If{$\tilde{q}_{k,l,kl}=\tilde{p}_{k,l}$ = 0}
                {$P_{k,l}$ $\leftarrow$ 0; $\tilde{LP}\leftarrow\tilde{LP}$ \textit{and} $\{\tilde{Q}\leftarrow\tilde{q}_{k,l,*,*}\leftarrow 0\}$; $\tilde{E} \leftarrow \tilde{E} \,\, \cup \,\, \{\tilde{p}_{k,l},\tilde{p}_{*,*}\}$\;
                }
           }
\eIf{Match(P)}
{\textbf{Top\_q}\;
\ForEach{$\{\tilde{p}_{u,i},\tilde{p}_{v,j}\}\in \mathcal{P}\backslash\tilde{E}\backslash\tilde{F}$}{
\nl    Maximize $\tilde{q}_{u,i,v,j}$ subject to $\tilde{LP}$\;
   \eIf{$Max<1$}
    {\textbf{Assign\_$\tilde{q}_{u,i,v,j}=0\equiv$}  $\{\tilde{LP}\leftarrow\tilde{LP}$ \textit{and} $\{\tilde{Q}\leftarrow\tilde{q}_{u,i,v,j}\leftarrow 0\}$; $\tilde{E}\leftarrow\tilde{E}\cup\{\tilde{p}_{u,i},\tilde{p}_{v,j}\}\}$\;    
    \textbf{goto} Top\_q\;}
{\nl Minimize $\tilde{q}_{u,i,v,j}$ subject to $\tilde{LP}$\;
\If{$Min>0$}
      {\textbf{Assign\_$\tilde{q}_{u,i,v,j}=1$}($u,i,v,j$,$\tilde{LP}$, $\tilde{E}$, $\tilde{Q}, \tilde{F}$)\;
\textbf{goto} Top\_p\;}
}
{$\tilde{q}_{u,i,v,j}$ attains 0 and 1 in Feasible Fractional Solutions of $\tilde{LP}$.\;}
}
All $\tilde{q}_{u,i,v,j}$ attain 0 and 1 in Feasible Fractional Solutions of $\tilde{LP}$.\;
RETURN Feasible Fractional Solution, $\tilde{LP}$, $\tilde{E}$, $\tilde{Q}$, $\tilde{F}$
}
{RETURN Infeasible, $\tilde{LP}$, $\tilde{E}$, $\tilde{Q}$, $\tilde{F}$}
}
\eIf{(true that lines 1 or 2 return `Infeasible' or `Feasible Integer Solution')}
   {RETURN Feasible Integer Solution, $\tilde{LP}$, $\tilde{E}$, $\tilde{Q}$, $\tilde{F}$}
   {RETURN Infeasible, $\tilde{LP}$, $\tilde{E}$, $\tilde{Q}$, $\tilde{F}$}\
\begin{center}{\textbf{Routine 1:} ImplementClosure}\end{center}
\label{alg:TedClosure} 
\end{algorithm}

\clearpage
\begin{algorithm}[h!]
\SetKwBlock{Block}{begin}{end}
\KwIn{($u,i,v,j$,$\tilde{\tilde{LP}}$, $\tilde{\tilde{E}}$, $\tilde{\tilde{Q}}$, $\tilde{\tilde{F}}$)} 
\KwOut{($\tilde{\tilde{LP}}$, $\tilde{\tilde{E}}$, $\tilde{\tilde{Q}}$, $\tilde{\tilde{F}}$)}

$\tilde{\tilde{LP}}\leftarrow\tilde{\tilde{LP}}$ \textit{and} $\tilde{\tilde{Q}}\leftarrow\tilde{\tilde{q}}_{u,i,v,j}\leftarrow 1$;  $\tilde{\tilde{F}}\leftarrow\tilde{\tilde{F}}\cup\{\tilde{\tilde{p}}_{u,i},\tilde{\tilde{p}}_{v,j}\}$\;
$\tilde{\tilde{LP}}\leftarrow\tilde{\tilde{LP}}$ \textit{and} $\tilde{\tilde{Q}}\leftarrow\tilde{\tilde{q}}_{u,i,u,i}=\tilde{\tilde{p}}_{u,i}\leftarrow 1$; $\tilde{\tilde{LP}}\leftarrow\tilde{\tilde{LP}}$ \textit{and} $\tilde{\tilde{Q}}\leftarrow\tilde{\tilde{q}}_{v,j,v,j}=\tilde{\tilde{p}}_{v,j}\leftarrow 1$\;

\Begin(`Zero' blocks in block rows $u$ \& $v$ in $\tilde{\tilde{Q}}$ EXCEPT blocks $[u,i]$ \& $[v,j]$.){
$\{\tilde{\tilde{LP}}\leftarrow\tilde{\tilde{LP}}$ \textit{and} $\{\tilde{\tilde{Q}}\leftarrow\tilde{\tilde{q}}_{u,l,*,*}\leftarrow 0 \,\,\&\,\, \tilde{\tilde{p}}_{u,l}\leftarrow 0\}$, $\tilde{\tilde{E}}\leftarrow\tilde{\tilde{E}}\cup\{\tilde{\tilde{p}}_{u,l},\tilde{\tilde{p}}_{*,*}\}$, $l=1,2,...,n$, $l\ne i\}$\;
$\{\tilde{\tilde{LP}}\leftarrow\tilde{\tilde{LP}}$ \textit{and} $\{\tilde{\tilde{Q}}\leftarrow\tilde{\tilde{q}}_{*,*,v,l}\leftarrow 0 \,\,\&\,\, \tilde{\tilde{p}}_{v,l}\leftarrow 0\}$, $\tilde{\tilde{E}}\leftarrow\tilde{\tilde{E}}\cup\{\tilde{\tilde{p}}_{*,*},\tilde{\tilde{p}}_{v,l}\}$, $l=1,2,...,n$, $l\ne j\}$\;
}
\Begin(`Zero' blocks in block columns $i$ \& $j$ in $\tilde{\tilde{Q}}$ EXCEPT blocks $[u,i]$ \& $[v,j]$.){
$\{\tilde{\tilde{LP}}\leftarrow\tilde{\tilde{LP}}$ \textit{and} $\{\tilde{\tilde{Q}}\leftarrow\tilde{\tilde{q}}_{l,i,*,*}\leftarrow 0 \,\,\&\,\, \tilde{\tilde{p}}_{l,i}\leftarrow 0\}$, $\tilde{\tilde{E}}\leftarrow\tilde{\tilde{E}}\cup\{\tilde{\tilde{p}}_{l,i},\tilde{\tilde{p}}_{*,*}\}$, $l=1,2,...,n$, $l\ne u\}$\;
$\{\tilde{\tilde{LP}}\leftarrow\tilde{\tilde{LP}}$ \textit{and} $\{\tilde{\tilde{Q}}\leftarrow\tilde{\tilde{q}}_{*,*,l,j}\leftarrow 0 \,\,\&\,\, \tilde{\tilde{p}}_{l,j}\leftarrow 0\}$, $\tilde{\tilde{E}}\leftarrow\tilde{\tilde{E}}\cup\{\tilde{\tilde{p}}_{*,*},\tilde{\tilde{p}}_{l,j}\}$, $l=1,2,...,n$, $l\ne v\}$\;
}
\Begin(`Zero rows $u$ \& $v$ in blocks $[u,i]$ \& $[v,j]$ EXCEPT components $[u,i]$ \& $[v,j]$.){
$\{\tilde{\tilde{LP}}\leftarrow\tilde{\tilde{LP}}$ \textit{and} $\{\tilde{\tilde{Q}}\leftarrow\tilde{\tilde{q}}_{u,l,v,j}\leftarrow 0$, $\tilde{\tilde{E}}\leftarrow\tilde{\tilde{E}}\cup\{\tilde{\tilde{p}}_{u,l},\tilde{\tilde{p}}_{v,j}\}$, $l=1,2,...,n$, $l\ne i\}$\;
$\{\tilde{\tilde{LP}}\leftarrow\tilde{\tilde{LP}}$ \textit{and} $\{\tilde{\tilde{Q}}\leftarrow\tilde{\tilde{q}}_{u,i,v,l}\leftarrow 0$, $\tilde{\tilde{E}}\leftarrow\tilde{\tilde{E}}\cup\{\tilde{\tilde{p}}_{u,i},\tilde{\tilde{p}}_{v,l}\}$, $l=1,2,...,n$, $l\ne j\}$\; 
}
\Begin(`Zero columns $i$ \& $j$ in blocks $[u,i]$ \& $[v,j]$ EXCEPT components $[u,i]$ \& $[v,j]$.){
$\{\tilde{\tilde{LP}}\leftarrow\tilde{\tilde{LP}}$ \textit{and} $\{\tilde{\tilde{Q}}\leftarrow\tilde{\tilde{q}}_{l,i,v,j}\leftarrow 0$, $\tilde{\tilde{E}}\leftarrow\tilde{\tilde{E}}\cup\{\tilde{\tilde{p}}_{l,i},\tilde{\tilde{p}}_{v,j}\}$, $l=1,2,...,n$, $l\ne u\}$\;
$\{\tilde{\tilde{LP}}\leftarrow\tilde{\tilde{LP}}$ \textit{and} $\{\tilde{\tilde{Q}}\leftarrow\tilde{\tilde{q}}_{u,i,l,j}\leftarrow 0$, $\tilde{\tilde{E}}\leftarrow\tilde{\tilde{E}}\cup\{\tilde{\tilde{p}}_{u,i},\tilde{\tilde{p}}_{l,j}\}$, $l=1,2,...,n$, $l\ne v\}$\;
}
RETURN $\tilde{\tilde{LP}}$, $\tilde{\tilde{E}}$, $\tilde{\tilde{Q}}$, $\tilde{\tilde{F}}$\
\begin{center}{\textbf{Routine 2:} Assume\_$\tilde{q}_{u,i,v,j}=1$}\end{center}
\label{alg:Assumeq1} 
\end{algorithm}

\clearpage
\begin{algorithm}[h!]
\SetKwBlock{Block}{begin}{end}
\KwIn{($\tilde{\tilde{LP}}$, $\tilde{\tilde{E}}$, $\tilde{\tilde{Q}}$, $\tilde{\tilde{F}}$)} 
\KwOut{(Decision, $\tilde{\tilde{LP}}$, $\tilde{\tilde{E}}$, $\tilde{\tilde{Q}}$, $\tilde{\tilde{F}}$)}
\While{(not true that lines 1 or 2 return `Infeasible' or `Feasible Integer Solution')}{
\textbf{Top\_p}\;
       \For{$k,l=1,2,...,n; $}{
            $P_{k,l}$ $\leftarrow$ 1\;
            \If{$\tilde{\tilde{q}}_{k,l,kl}=\tilde{\tilde{p}}_{k,l}$ = 0}
                {$P_{k,l}$ $\leftarrow$ 0; $\tilde{\tilde{LP}}\leftarrow\tilde{\tilde{LP}}$ \textit{and} $\{\tilde{\tilde{Q}}\leftarrow\tilde{\tilde{q}}_{k,l,*,*}\leftarrow 0\}$; $\tilde{\tilde{E}} \leftarrow \tilde{\tilde{E}} \,\, \cup \,\, \{\tilde{\tilde{p}}_{k,l},\tilde{\tilde{p}}_{*,*}\}$\;
                }
           }
\eIf{Match(P)}
{\textbf{Top\_q}\;
\ForEach{$\{\tilde{\tilde{p}}_{u,i},\tilde{\tilde{p}}_{v,j}\}\in \mathcal{P}\backslash\tilde{\tilde{E}}\backslash\tilde{\tilde{F}}$}{
\nl    Maximize $\tilde{\tilde{q}}_{u,i,v,j}$ subject to $\tilde{\tilde{LP}}$\;
   \eIf{$Max<1$}
    {\textbf{Assign\_$\tilde{\tilde{q}}_{u,i,v,j}=0\equiv$}  $\{\tilde{\tilde{LP}}\leftarrow\tilde{\tilde{LP}}$ \textit{and} $\{\tilde{\tilde{Q}}\leftarrow\tilde{\tilde{q}}_{u,i,v,j}\leftarrow 0\}$; $\tilde{\tilde{E}}\leftarrow\tilde{\tilde{E}}\cup\{\tilde{\tilde{p}}_{u,i},\tilde{\tilde{p}}_{v,j}\}\}$\;    
    \textbf{goto} Top\_q\;}
{\nl Minimize $\tilde{\tilde{q}}_{u,i,v,j}$ subject to $\tilde{\tilde{LP}}$\;
\If{$Min>0$}
      {\textbf{Assign\_$\tilde{\tilde{q}}_{u,i,v,j}=1$}($u,i,v,j$,$\tilde{\tilde{LP}}$, $\tilde{\tilde{E}}$, $\tilde{\tilde{Q}}, \tilde{\tilde{F}}$)\;
\textbf{goto} Top\_p\;}
}
{$\tilde{\tilde{q}}_{u,i,v,j}$ attains 0 and 1 in Feasible Fractional Solutions of $\tilde{\tilde{LP}}$.\;}
}
All $\tilde{\tilde{q}}_{u,i,v,j}$ attain 0 and 1 in Feasible Fractional Solutions of $\tilde{\tilde{LP}}$.\;
RETURN Feasible Fractional Solution, $\tilde{\tilde{LP}}$, $\tilde{\tilde{E}}$, $\tilde{\tilde{Q}}$, $\tilde{\tilde{F}}$
}
{RETURN Infeasible, $\tilde{\tilde{LP}}$, $\tilde{\tilde{E}}$, $\tilde{\tilde{Q}}$, $\tilde{\tilde{F}}$}
}
\eIf{(true that lines 1 or 2 return `Infeasible' or `Feasible Integer Solution')}
   {RETURN Feasible Integer Solution, $\tilde{\tilde{LP}}$, $\tilde{\tilde{E}}$, $\tilde{\tilde{Q}}$, $\tilde{\tilde{F}}$}
   {RETURN Infeasible, $\tilde{\tilde{LP}}$, $\tilde{\tilde{E}}$, $\tilde{\tilde{Q}}$, $\tilde{\tilde{F}}$}\
\begin{center}{\textbf{Routine 3:} TestAssumption}\end{center}   
\label{alg:TestAssumption} 
\end{algorithm}

\clearpage
\begin{algorithm}[h!]
\SetKwBlock{Block}{begin}{end}
\KwIn{($u,i,v,j$,$\tilde{LP}$, $\tilde{E}$, $\tilde{Q}$, $\tilde{F}$)} 
\KwOut{($\tilde{LP}$, $\tilde{E}$, $\tilde{Q}$, $\tilde{F}$)}

$\tilde{LP}\leftarrow\tilde{LP}$ \textit{and} $\tilde{Q}\leftarrow\tilde{q}_{u,i,v,j}\leftarrow 1$;  $\tilde{F}\leftarrow\tilde{F}\cup\{\tilde{p}_{u,i},\tilde{p}_{v,j}\}$\;
$\tilde{LP}\leftarrow\tilde{LP}$ \textit{and} $\tilde{Q}\leftarrow\tilde{q}_{u,i,u,i}=\tilde{p}_{u,i}\leftarrow 1$; $\tilde{LP}\leftarrow\tilde{LP}$ \textit{and} $\tilde{Q}\leftarrow\tilde{q}_{v,j,v,j}=\tilde{p}_{v,j}\leftarrow 1$\;
\Begin(`Zero' blocks in block rows $u$ \& $v$ in $\tilde{Q}$ EXCEPT blocks $[u,i]$ \& $[v,j]$.){
$\{\tilde{LP}\leftarrow\tilde{LP}$ \textit{and} $\{\tilde{Q}\leftarrow\tilde{q}_{u,l,*,*}\leftarrow 0 \,\,\&\,\, \tilde{p}_{u,l}\leftarrow 0\}$, $\tilde{E}\leftarrow\tilde{E}\cup\{\tilde{p}_{u,l},\tilde{p}_{*,*}\}$, $l=1,2,...,n$, $l\ne i\}$\;
$\{\tilde{LP}\leftarrow\tilde{LP}$ \textit{and} $\{\tilde{Q}\leftarrow\tilde{q}_{*,*,v,l}\leftarrow 0 \,\,\&\,\, \tilde{p}_{v,l}\leftarrow 0\}$, $\tilde{E}\leftarrow\tilde{E}\cup\{\tilde{p}_{*,*},\tilde{p}_{v,l}\}$, $l=1,2,...,n$, $l\ne j\}$\;
}
\Begin(`Zero' blocks in block columns $i$ \& $j$ in $\tilde{Q}$ EXCEPT blocks $[u,i]$ \& $[v,j]$.){
$\{\tilde{LP}\leftarrow\tilde{LP}$ \textit{and} $\{\tilde{Q}\leftarrow\tilde{q}_{l,i,*,*}\leftarrow 0 \,\,\&\,\, \tilde{p}_{l,i}\leftarrow 0\}$, $\tilde{E}\leftarrow\tilde{E}\cup\{\tilde{p}_{l,i},\tilde{p}_{*,*}\}$, $l=1,2,...,n$, $l\ne u\}$\;
$\{\tilde{LP}\leftarrow\tilde{LP}$ \textit{and} $\{\tilde{Q}\leftarrow\tilde{q}_{*,*,l,j}\leftarrow 0 \,\,\&\,\, \tilde{p}_{l,j}\leftarrow 0\}$, $\tilde{E}\leftarrow\tilde{E}\cup\{\tilde{p}_{*,*},\tilde{p}_{l,j}\}$, $l=1,2,...,n$, $l\ne v\}$\;
}
\Begin(`Zero rows $u$ \& $v$ in blocks $[u,i]$ \& $[v,j]$ EXCEPT components $[u,i]$ \& $[v,j]$.){
$\{\tilde{LP}\leftarrow\tilde{LP}$ \textit{and} $\{\tilde{Q}\leftarrow\tilde{q}_{u,l,v,j}\leftarrow 0$, $\tilde{E}\leftarrow\tilde{E}\cup\{\tilde{p}_{u,l},\tilde{p}_{v,j}\}$, $l=1,2,...,n$, $l\ne i\}$\;
$\{\tilde{LP}\leftarrow\tilde{LP}$ \textit{and} $\{\tilde{Q}\leftarrow\tilde{q}_{u,i,v,l}\leftarrow 0$, $\tilde{E}\leftarrow\tilde{E}\cup\{\tilde{p}_{u,i},\tilde{p}_{v,l}\}$, $l=1,2,...,n$, $l\ne j\}$\; 
}
\Begin(`Zero columns $i$ \& $j$ in blocks $[u,i]$ \& $[v,j]$ EXCEPT components $[u,i]$ \& $[v,j]$.){
$\{\tilde{LP}\leftarrow\tilde{LP}$ \textit{and} $\{\tilde{Q}\leftarrow\tilde{q}_{l,i,v,j}\leftarrow 0$, $\tilde{E}\leftarrow\tilde{E}\cup\{\tilde{p}_{l,i},\tilde{p}_{v,j}\}$, $l=1,2,...,n$, $l\ne u\}$\;
$\{\tilde{LP}\leftarrow\tilde{LP}$ \textit{and} $\{\tilde{Q}\leftarrow\tilde{q}_{u,i,l,j}\leftarrow 0$, $\tilde{E}\leftarrow\tilde{E}\cup\{\tilde{p}_{u,i},\tilde{p}_{l,j}\}$, $l=1,2,...,n$, $l\ne v\}$\;
}
RETURN $\tilde{LP}$, $\tilde{E}$, $\tilde{Q}$, $\tilde{F}$\

\begin{center}{\textbf{Routine 4:} Assign\_$\tilde{q}_{u,i,v,j}=1$}\end{center}
\label{alg:Assignq1} 
\end{algorithm}

\bibliographystyle{siam.bst}

\end{document}